\documentclass[a4paper,11pt]{article} 
\usepackage{amssymb,amsmath,amsfonts,makeidx,placeins,pbox,multirow}
\usepackage{rotate,color,slashed,cite,caption,epstopdf,verbatim,url,multirow}
\usepackage{epstopdf}
\usepackage{subfig}
\usepackage{graphicx}
\usepackage{longtable,tabu}  \usepackage{array}
\newcolumntype{P}[1]{>{\centering\arraybackslash}p{#1}}
\usepackage[colorlinks=true, linkcolor=magenta, urlcolor=blue,
citecolor=blue]{hyperref} \usepackage[utf8x]{inputenc}

\numberwithin{equation}{section}

\long\def\change#1!!!#2!!!{\color{red}#1 \color{blue}#2 \color{black}}

\evensidemargin=\oddsidemargin \topmargin -1.0cm 
\begin{document}
	
\begin{center}
		
{\Large \bf Probing composite Higgs boson substructure at the HL-LHC} \\
\vspace*{0.5cm} {\sf Avik Banerjee~$^{a,b,}$\footnote{avik@chalmers.se}, ~Sayan Dasgupta~$^{c,}$\footnote{sayandg05@gmail.com}, ~Tirtha Sankar Ray~$^{c,}$\footnote{tirthasankar.ray@gmail.com}} \\
\vspace{10pt} {\small } $^{a)}$ {\em Department of Physics, Chalmers University of Technology, Fysikg\aa rden, 41296 G\"oteborg, Sweden}
\\
\vspace{3pt} {\small } $^{b)}${\em Saha Institute of Nuclear Physics, HBNI, 1/AF Bidhan Nagar, Kolkata 700064, India}
\\
\vspace{3pt} {\small } $^{c)}${\em Department of Physics, Indian Institute of Technology Kharagpur, Kharagpur 721302, India} 
\normalsize
\end{center}

\date{}
	

\begin{abstract}

The Higgs boson may well be a composite scalar with a finite extension in space. Owing to the momentum dependence of its couplings the imprints of such a composite pseudo Goldstone Higgs may show up in the tails of various kinematic distributions at the LHC, distinguishing it from an elementary state. From the bottom up we construct the momentum dependent form factors to capture the interactions of the composite Higgs with the weak gauge bosons. We demonstrate their impact in the differential distributions of various kinematic parameters for the $pp\rightarrow Z^*H\rightarrow l^+l^-b\bar{b}$ channel. We show that this channel can provide an important handle to probe the Higgs' substructure at the HL-LHC.
	
\end{abstract}


\bigskip


\section{Introduction}

Since its discovery \cite{Aad:2012tfa,Chatrchyan:2012xdj} the properties of the Higgs particle have been under intense theoretical and experimental scrutiny. A primary question relates to the existence of possible internal structure of the Higgs boson. The upcoming runs of the Large Hadron Collider (LHC) and all future collider experiments are mandated to study the properties of the Higgs including exploring the possibility of the Higgs having a finite extension in space.

Interestingly a composite Higgs having a non-trivial internal structure is known to be a handle in addressing the notorious gauge hierarchy problem of the Standard Model (SM) \cite{Kaplan:1983fs,Dugan:1984hq,Contino:2003ve,Agashe:2004rs,Contino:2010rs,Panico:2015jxa,Erdmenger:2020lvq,Erdmenger:2020flu}. In this paper we consider the well-motivated composite Higgs framework  where the Higgs is identified with a pseudo Nambu-Goldstone boson (pNGB) of a strongly interacting sector. Within this paradigm the Higgs boson may be considered as a composite of underlying chiral fermions \cite{Barnard:2013zea,Ferretti:2013kya,Ferretti:2014qta}. The compositeness scale $\Lambda_H$ is related to the length scale $l_H\sim 1/\Lambda_H$ governing the geometric size of the Higgs. To conform with the electroweak precision data a separation of $\Lambda_H$ and the weak scale ($v<<\Lambda_H$) is introduced by considering the Higgs as a pNGB of the strong sector. In this framework existence of new resonances near the compositeness scale ($m_Q\sim \Lambda_H$) is expected, which can be searched directly at the LHC \cite{Barcelo:2011wu, DeSimone:2012fs, Azatov:2015xqa, Moretti:2016gkr, Cacciapaglia:2018qep, Dasgupta:2019yjm,  Low:2015uha,Niehoff:2015iaa,Franzosi:2016aoo,Liu:2018hum}. These exotic states along with the modification of Higgs couplings are the two major conventional signatures of this setup that have been actively chased in collider experiments \cite{Falkowski:2007hz,Banerjee:2017wmg,Banerjee:2020tqc,Carragher:2021qaj,Khosa:2021wsu,Cao:2018cms,Xie:2021xtl,Yan:2021veo}. The modifications of the Higgs couplings originate from two sources. First, a deviation from the SM value of the coupling arises due to the non-linear structure of the pNGB chiral Lagrangian. The measurements of the Higgs signal strengths at the LHC constrain these deviations to less than 10\%-15\% from their SM values \cite{Falkowski:2007hz,Banerjee:2017wmg,Banerjee:2020tqc,Carragher:2021qaj,Khosa:2021wsu}. An additional source of modification arises from the extension of the Higgs in space resulting in a dramatic scaling of the Higgs coupling with the transferred momentum. It is this latter phenomenon that provides a more direct evidence of the non-elementary nature of the Higgs \cite{ Azatov:2013xha, Azatov:2014jga, Murayama:2014yja,Bellazzini:2015cgj,  Azatov:2016xik, Goncalves:2018pkt, BuarqueFranzosi:2019dwg, Goncalves:2020vyn,mastersthesis}.

We construct the Higgs-elementary coupling form factors from the bottom-up, relying on the empirical Lorentz structure and inspiration from large $N$ tools for composite states \cite{tHooft:1973alw,tHooft:1974pnl,Witten:1979kh}, assuming an underlying strongly interacting hypercolor dynamics. We demonstrate that the high luminosity LHC (HL-LHC) \cite{Cepeda:2019klc} can explore a significant portion of the parameter space where the momentum dependence of the Higgs coupling implies pronounced deviation from the SM predictions in the differential distribution of kinematic parameters. This provides an interesting handle to explore the Higgs compositeness, even if the compositeness scale lies just beyond the LHC reach. As a proof of principle, we consider the $pp\rightarrow Z^*H\rightarrow l^+l^-b\bar{b}$ channel at the HL-LHC and demonstrate that there is a considerable deviation in the kinematic distributions of the final state objects for this channel that can be used to explore the non-elementary nature of the Higgs boson.

The rest of the paper is organized as follows. In Section~\ref{sec:theory} we construct the form factors associated to the Higgs couplings with the weak gauge bosons.
In Section~\ref{sec:collider} we perform the collider analysis to demonstrate the importance of the differential distributions in probing the composite nature of the Higgs boson. In this regard, we show the prospects of the HL-LHC in Section~\ref{sec:HLLHC}, before concluding in Section~\ref{sec:conclusion}.   

\section{Composite Higgs couplings}
\label{sec:theory}

In the SM, the Higgs couplings to the massive gauge bosons ($V=W^\pm,~Z$) are given by
\begin{equation}
\label{hvv_SM}
\mathcal{L}_{hVV}=g^{\rm SM}_V h V_\mu V^\mu\,, \quad {\rm where} \quad \left\{g^{\rm SM}_W\,,~g^{\rm SM}_Z\right\}=\left\{\frac{g^2v}{2}\,,\frac{g^2v}{2\cos^2\theta_W}~\right\}\,, \quad {\rm and} \quad v=246~{\rm GeV}\,.
\end{equation}
On the other hand, in composite Higgs frameworks the above coupling can in general be written as
\begin{equation}
\label{hvv_FF_1}
\mathcal{L}_{hVV}=g^{\rm SM}_V\frac{\Pi^{\mu\nu}_V(p_1,p_2)}{f^2} h V_\mu(p_1) V_\nu(p_2)\,,
\end{equation}
where the scale $f$ denotes the decay constant of the pNGB Higgs. The compositeness scale $\Lambda_H$ is related to $f$ as $\Lambda_H\sim g_*f$, where $g_*$ represents a generic strong sector coupling as $1\ll g_*<4\pi$. 
The momentum dependent form factor $\Pi^{\mu\nu}_V$ captures the non-perturbative dynamics of the strong sector leading to a non-elementary Higgs with a finite shape. Assuming the Higgs to be a purely CP-even state, the Lorentz structure of the form factor can be decomposed as \cite{Isidori:2013cla,Isidori:2013cga,Bellazzini:2015cgj}
\begin{equation}
\label{hvv_FF_2}
\Pi^{\mu\nu}_{V}=\left[\Pi_1^V \eta^{\mu\nu} + \frac{1}{\Lambda_H^2}\big\{ \Pi_2^V \left(\eta^{\mu\nu}p_1.p_2-p_2^\mu p_1^\nu\right) + \Pi_3^V p_1^\mu p_2^\nu  + \Pi_4^V p_1^\mu p_1^\nu + \Pi_5^V p_2^\mu p_2^\nu \big\}\right]\,. 
\end{equation}
In the above equation, the momentum-dependent functions $\Pi_i^V(p_1,p_2)$ have mass dimension 2, and we have extracted appropriate powers of $\Lambda_H$ on the basis of dimensional analysis. Demanding that the SM coupling is reproduced at the low energy ($p_i.p_j\ll \Lambda_H^2$), \textit{modulo} a suppression due to the nonlinearity of the pNGB, we can constrain the IR behaviour of $\Pi^{\mu\nu}_V$ as
\begin{equation}
\label{low_en}
\lim_{p_i.p_j\ll\Lambda_H^2}g^{\rm SM}_V\frac{\Pi^{\mu\nu}_V(p_1,p_2)}{f^2}\simeq g^{\rm SM}_V\frac{\Pi_1^V(0,0)}{f^2}\eta^{\mu\nu}= g^{\rm SM}_V\sqrt{1-\xi}\eta^{\mu\nu}\,.
\end{equation}

The usual suppression by a factor of $\sqrt{1-\xi}$ (where $\xi\equiv v^2/f^2$) is considered to reproduce the $hVV$ coupling in the minimal composite Higgs model. A full non-perturbative calculation is in general required to find the detailed momentum dependence of the form factors. However, one can guess some ansatz for $\Pi_i^V$ from the wisdom of large $N$ formalism \cite{Agashe:2004rs,Contino:2010rs,Pomarol:2012qf,Marzocca:2012zn,Orgogozo:2011kq}. We follow this approach below to adopt an ansatz for the three-point vertex form factors.

\subsection{Large $\bf N$ implications}

\begin{figure}[t]
	\centering
	\subfloat[\label{fig_correl_a}]{\includegraphics[trim = 20mm 220.0mm 130mm 25mm, clip,width=0.3\textwidth, scale=0.2]{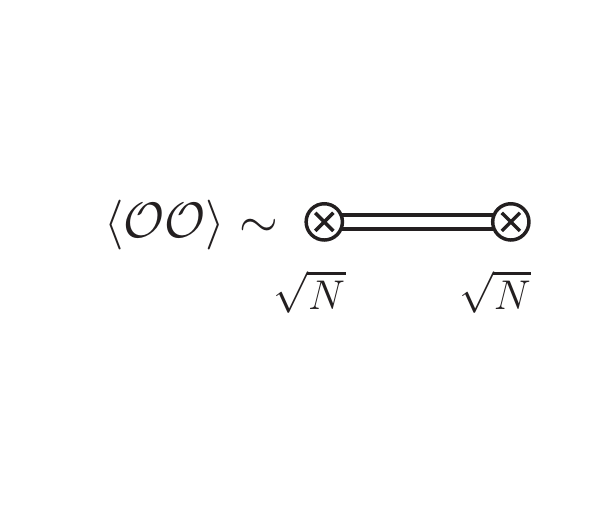}}
	\hspace{0.1cm}
	\subfloat[\label{fig_correl_b}]{\includegraphics[trim = 80mm 220.0mm 70mm 25mm, clip,width=0.3\textwidth, scale=0.2]{FF_largeN.pdf}}
	\hspace{0.1cm}
	\subfloat[\label{fig_correl_c}]{\includegraphics[trim = 140mm 220.0mm 10mm 25mm, clip,width=0.3\textwidth, scale=0.2]{FF_largeN.pdf}}
	\caption{\small\it Large $N$ scaling for two-point (left) and three-point (right) correlators in a strongly coupled SU(N) gauge theory.}
	\label{fig_correl}
\end{figure}

To progress further we assume that the  strong sector has an underlying confining $\rm SU(N)$ gauge dynamics with a suitably large $N$. In the large $N$ limit, the $n$-point correlators between composite operators $\mathcal{O}_1, \mathcal{O}_2,...\mathcal{O}_n$ of this confining theory scales as $\langle \mathcal{O}_1 \mathcal{O}_2...\mathcal{O}_n \rangle\sim N$ at the leading order \cite{tHooft:1973alw,tHooft:1974pnl,Witten:1979kh}. In this limit one can show that a two-point correlation function as shown in Fig.\,\ref{fig_correl_a} can be approximated at the leading order in $N$ as
\begin{equation}
\langle \mathcal{O}(p)\mathcal{O}(-p)\rangle \sim \sum_{a=1}^{\infty} \frac{F_a^2}{p^2-m_a^2+i{\rm Im}[M^2_a(p)]}\,, \quad F_a\equiv \langle 0|\mathcal{O}|a\rangle\,.
\end{equation} 
The masses of the single-particle mesonic states are given as $m_a\equiv g_aF_a$, where $g_a\sim 1/\sqrt{N}$. The large $N$ counting shows that the decay constant $F_a\sim \sqrt{N}$, so that $\langle\mathcal{O}\mathcal{O}\rangle\sim N$. Here $M^2_a(p)$ represents the radiative corrections to the 1PI resummed propagators of the mesonic states. Similar arguments can be extended to three-point correlators as well, where two different structures are possible as follows \cite{Witten:1979kh} (see Figs.\,\ref{fig_correl_b} and \ref{fig_correl_c}):
\begin{equation}
\label{three_pt_1}
\langle \mathcal{O}(p_1)\mathcal{O}(p_2)\mathcal{O}(-p_1-p_2)\rangle \sim \sum_{a,b=1}^{\infty} \frac{F_a F_b \langle 0|\mathcal{O}|ab\rangle }{\left(p_1^2-m_a^2+i{\rm Im}[M^2_a(p)]\right)\left(p_2^2-m_b^2+i{\rm Im}[M^2_b(p)]\right)}\,.
\end{equation}
and,
\begin{align}
\label{three_pt_2}
\nonumber
& \langle \mathcal{O}(p_1)\mathcal{O}(p_2)\mathcal{O}(-p_1-p_2)\rangle \sim \\
& \sum_{a,b,c=1}^{\infty}\frac{F_a F_bF_c\Gamma_{c\to ab}}{\left(p_1^2-m_a^2+i{\rm Im}[M^2_a(p)]\right)\left(p_2^2-m_b^2+i{\rm Im}[M^2_b(p)]\right)\left((p_1+p_2)^2-m_c^2+i{\rm Im}[M^2_c(p)]\right)}\,.
\end{align}
In the first case, one of the operators creates two meson-states (with amplitude $\langle 0|\mathcal{O}|ab\rangle$) which are annihilated by the other two operators, while in the second case each operator excites a single meson and they interact via a local three-point vertex (given by $\Gamma_{c\to ab}$). Clearly, the leading $N$ scaling suggests that $\langle 0|\mathcal{O}|ab\rangle\sim1$ and $\Gamma_{c\to ab}\sim 1/\sqrt{N}$. 

We apply these results to the composite pNGB Higgs scenario. The relevant correlation function for the $hVV$ coupling is a correlator between three currents $\langle J_\mu J_\nu J_\pi \rangle$. 
The vector currents $J_{\mu, \nu }$ mix with the weak elementary gauge bosons implying a linear mixing between the $W^\pm$ and $Z$ bosons with composite spin-1 mesons. The current $J_\pi$ can excite a pNGB Higgs boson which is a purely composite condensate. In Fig.\,\ref{fig_correl_b}, the Higgs boson is generated from the vacuum by non-perturbative dynamics with a strength proportional to $\langle 0|J_\pi|ab\rangle$, where $a,b$ denotes the two spin-1 meson states. On the other hand in Fig.\,\ref{fig_correl_c}, the pNGB Higgs which is denoted by the horizontal double line, interacts with the composite mesons through derivative couplings. As a result, the coupling strength $\Gamma_{c\to ab}$ in Eq.\,\eqref{three_pt_2} has an additional $\mathcal{O}(p_h^2/\Lambda_H^2)$ suppression in comparison to Eq.\,\eqref{three_pt_1}, where $p_h$ denotes the momentum passing through the Higgs\footnote{Another possibility, not considered in this paper, may arise if an elementary Higgs boson mixes with a composite scalar \cite{Galloway:2016fuo, Agugliaro:2016clv, Alanne:2017rrs}. In that case $\Gamma_{c\to ab}$ denotes local interaction between the additional composite scalar with two spin-1 mesons, which may arise at the same order as Eq.\,\eqref{three_pt_1}.}. It is expected that the underlying strong dynamics leads to a unique meson spectrum. As a consequence all the form factors $\Pi_i^V$ should have identical pole structure in the large $N$ limit.
Equipped with these results
we adopt an ansatz for $\Pi_i^V$ at the leading order in $N$ as well as in the momentum passing through the Higgs as
\begin{equation}
\label{gamma_expr_1}
\Pi_i(p_1,p_2)=\sum_{a,b=1}^{\infty}\frac{c^V_i F_a F_b \langle 0|J_\pi|ab\rangle}{\left(p_1^2-m_a^2+i{\rm Im}[M^2_a(p)]\right)\left(p_2^2-m_b^2+i{\rm Im}[M^2_b(p)]\right)}\,.
\end{equation}
The constant coefficients $c^V_i$ parametrizes our ignorance about the details of the strong dynamics. Although the form factor contains an infinite sum over the meson-states, the convergence of the series is ensured by assuming that the masses of the mesonic states appearing in the successive terms are hierarchical  \cite{Agashe:2004rs,Contino:2010rs,Pomarol:2012qf,Marzocca:2012zn,Orgogozo:2011kq}.  
We take the dominant contribution coming from the first term in the summation and use the low energy constraint given in Eq.\,\eqref{low_en}, to obtain
\begin{align}
\label{hvv_FF_3}
\nonumber
\Pi^{\mu\nu}_{V}=\frac{f^2m_1^2m_2^2\left(1-i\frac{\Gamma_1}{m_1}\right)\left(1-i\frac{\Gamma_2}{m_2}\right)}{(p_1^2-m_1^2+im_1\Gamma_1)(p_2^2-m_2^2+im_2\Gamma_2)} \left[\sqrt{1-\xi}\eta^{\mu\nu} + \frac{1}{m_1m_2}\big\{ c^V_2\left(\eta^{\mu\nu}p_1.p_2-p_2^\mu p_1^\nu\right) \right.\\ 
+ c^V_3 p_1^\mu p_2^\nu  + c^V_4 p_1^\mu p_1^\nu + c^V_5 p_2^\mu p_2^\nu \big\}\bigg]\,. 
\end{align}
In the above equation, we use the narrow width approximation to replace ${\rm Im}[M^2_a(p)]$ by $m_a\Gamma_a$, where $\Gamma_a$ denotes the total width of the mesonic states and identified $\Lambda_H\equiv\sqrt{m_1m_2}$. 
Interestingly, the insertion of two lightest meson states in the form factor as above provides a handle to arrive at a convergent radiative Coleman-Weinberg potential for the pNGB Higgs \cite{Contino:2010rs,Pomarol:2012qf,Marzocca:2012zn}.

The strongly interacting light Higgs (SILH) description provides an alternative way to describe the composite Higgs models through an effective field theory framework \cite{Giudice:2007fh,Alloul:2013naa,Maltoni:2013sma,Ellis:2014dva,Biekotter:2014gup,Mimasu:2015nqa,Brehmer:2015rna,Degrande:2016dqg,Englert:2016hvy}. The relevant custodial $hVV$ operators at leading order are given by 
\begin{align}
\nonumber
\mathcal{L}_{\rm SILH}&\supset\frac{c_H}{2f^2}\partial^\mu(H^\dagger H)\partial_\mu(H^\dagger H)+\frac{i c_W g}{2m_\rho^2}(H^\dagger\sigma^a\overleftrightarrow{D^\mu}H)D^\nu W^a_{\mu\nu}+\frac{i c_B g^\prime}{2m_\rho^2}(H^\dagger\overleftrightarrow{D^\mu}H)\partial^\nu B_{\mu\nu}\\
\label{silh}
&+\frac{i c_{HW} g}{16\pi^2f^2}(D^\mu H^\dagger)\sigma^a(D^\nu H)W^a_{\mu\nu}+\frac{i c_{HB} g^\prime}{16\pi^2f^2}(D^\mu H^\dagger)(D^\nu H)B_{\mu\nu}\,.
\end{align}
If we identify $m_1=m_2=m_\rho\lesssim 4\pi f$, then the effective $hVV$ coupling obtained from the SILH Lagrangian can be mapped into the leading order expansion $\sim\mathcal{O}(p^2/m_\rho^2)$  of the form factor as shown in Appendix\,\ref{SILH_map}.

We take a quick look at the existing constraints on the $hVV$ couplings to device our benchmark. Modification of the Higgs couplings by measuring inclusive cross sections of the Higgs production processes at the LHC Run 2 exclude $f\leq 1$ TeV \cite{Banerjee:2017wmg,Banerjee:2020tqc}. Although the constraints from electroweak precision data have considerable dependence on the UV constructions \cite{Contino:2010rs,Agashe:2005dk,Orgogozo:2011kq,Falkowski:2013dza,Contino:2015mha}, a somewhat conservative limit of $m_{1,2}\gtrsim 2 - 3$ TeV, is reported in \cite{Ghosh:2015wiz}. Similarly direct search of non-Standard vector bosons at the LHC puts a limit on the masses of these exotic states in the same ballpark region  \cite{Aaboud:2017cxo, Low:2015uha,Niehoff:2015iaa,Franzosi:2016aoo,Liu:2018hum}, though the individual analyses involve several assumptions on model parameters and branching ratios to specific channels. Considering the values of $\xi$, allowed by the Higgs coupling measurements, we find that the limits on the $m_{1,2}$ from unitarity \cite{Bellazzini:2012tv,BuarqueFranzosi:2017prc} is considerably weaker than the electroweak precision data and direct search limits. 
Current limits on the SILH coefficients are extracted from \cite{Ellis:2018gqa,ATLAS:2019yhn,CMS:2020gsy,Ethier:2021bye} and listed in the Appendix\,\ref{SILH_map}. 
However, using these limits to arbitrarily high momenta ($p^2\gg M_W^2$) is not justified as it will violate the EFT expansion \cite{Isidori:2013cga}.
\begin{figure}[t!]
	\centering
	{\includegraphics[trim={2cm 20cm 6cm 3.9cm},clip,scale=0.85]{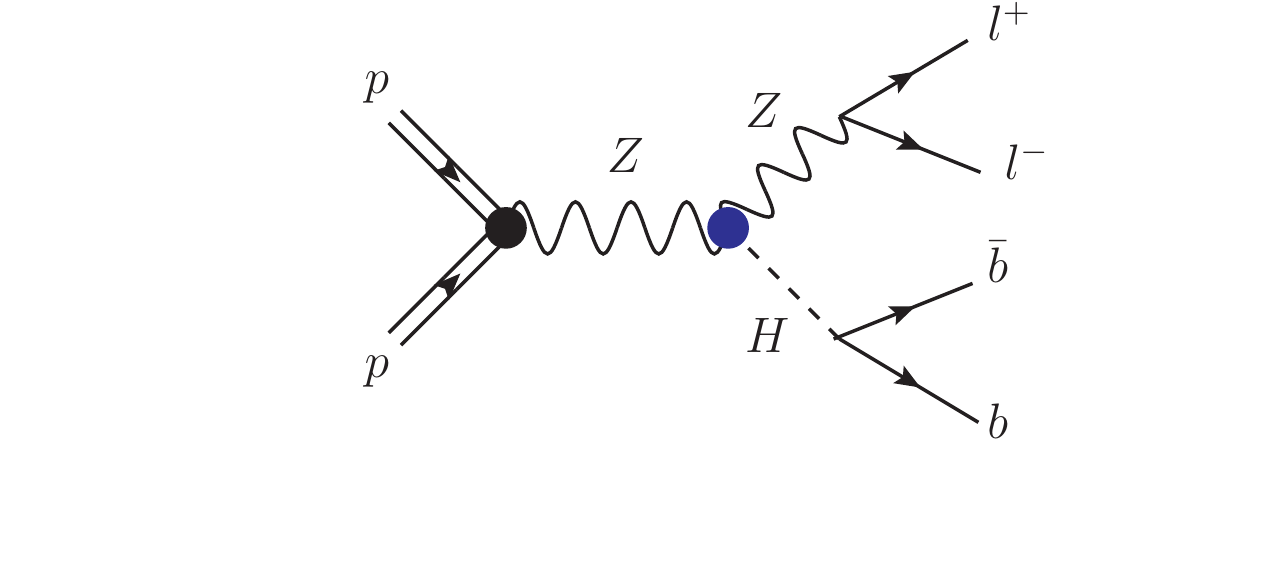}}
	\caption{\small\it Feynman diagram corresponding to the $pp\to Z^*H\to l^+l^-b\bar{b}$ process. The form factor involved in the $hZZ$ vertex is denoted by the blue blob.}
	\label{fig:feynman_zh}
\end{figure}

\section{Probing the Higgs form factor at collider experiments}
\label{sec:collider}

In this section we explore the possibility of utilizing the $pp\to Z^*H\to l^+l^-b\bar{b}$ channel (where $l=e,\mu$, see Fig.\,\ref{fig:feynman_zh}) to probe the $hZZ$ form factor introduced in Eqs.~\eqref{hvv_FF_1} and \eqref{hvv_FF_3}. In contrast to the conventional searches for the Higgs-strahlung process at the LHC  \cite{Sirunyan:2018kst,Aaboud:2018zhk}, we focus on the tails of the distributions of various kinematic observables where the $Z$ boson is far off-shell. Despite a relatively low cross-section it is easier to get an off-shell $Z$ in the associated Higgs production mode in comparison to the weak boson fusion or processes involving decay of Higgs into diboson. In particular, the presence of a $s$-channel $Z$ boson in the $Z^*H$ production mode proves to be most advantageous to probe the impact of the form factor. 

The form factor is implemented in the SM Universal Feynrules Output (UFO) format in {\tt MadGraph5} \cite{Alwall:2014hca}. Guided by the existing constraints presented in the previous section we consider $\xi=0.06$, $m_1=m_2=2.5$ TeV and $c_{i\ne 1}^{Z}=1$ as the benchmark values to present our results. Note that the second term in Eq.\,\eqref{hvv_FF_2} also receives a contribution from the SM at one loop, however, we have encoded that contribution inside the free parameter $c^Z_2$ together with the effects from the strong dynamics. 
Composite Higgs models generically predict the existence of broad spin-1 resonances \cite{Liu:2019bua,Jung:2019iii}.
Here we take a purely phenomenological approach and choose two extreme benchmark cases $\Gamma_{1,2}/m_{1,2}=1\%$ and $20\%$ to compare the impact of a narrow and a broad resonance in the differential distributions. We assume SM-like Higgs-bottom coupling in our analysis\footnote{Unlike the $hVV$ term, the modification of the Yukawa coupling in partial compositeness framework depends on the details of the model. More specifically it depends on which representation of the global symmetry of the strong sector is used to embed the bottom quark. To avoid this model specific uncertainties, we assume the $hb\bar{b}$ coupling to be equal to its SM value. As an example, in MCHM$_4$ scenario modification of the $hb\bar{b}$ coupling leads to a overall reduction of the cross-section by 6\% (for $f=1$ TeV).}.
\begin{figure}[t!]
	\centering
	{\includegraphics[scale=0.35]{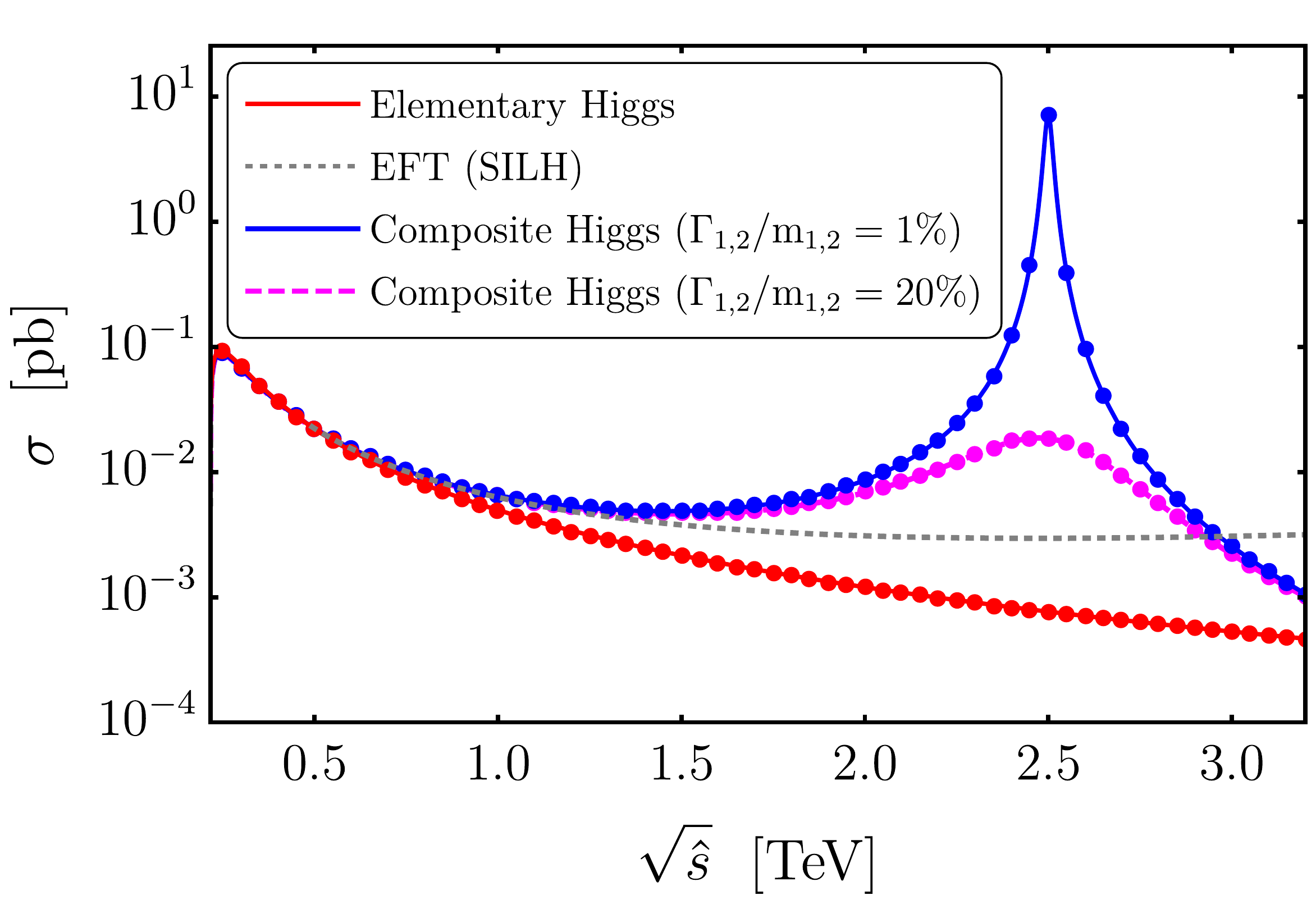}}
	\caption{\small\it Variation of cross section given in Eqs.~\eqref{analytic} and \eqref{analytic_2} with partonic centre of mass energy ($\sqrt{\hat{s}}$) for the elementary (red) and the composite Higgs scenario with $\Gamma_{1,2}/m_{1,2}=1\%$ (blue) and $20\%$ (magenta, dashed) are displayed. For comparison, the cross section for the SILH framework (grey, dotted) is also shown. To validate the implementation of the form factor in {\tt MadGraph5}, we plot the simulated cross section by different colored points.}
	\label{fig:zh_parton}
	\label{fig:partonic}
\end{figure} 

The parton level cross section for the Higgs-strahlung process ($q\bar{q}\to ZH$), assuming $m_1=m_2$ ($\Gamma_1=\Gamma_2$) and neglecting the light quark masses is given by
\begin{equation}
\small
\label{analytic}
\sigma(\hat{s})=\frac{\sigma^{\rm SM}(\hat{s})m_1^8\left(1+\frac{\Gamma_1^2}{m_1^2}\right)^2}{\left[(\hat{s}-m_1^2)^2+m_1^2\Gamma_1^2\right]\left[(M_Z^2-m_1^2)^2+m_1^2\Gamma_1^2\right]}\left[(1-\xi)-\frac{24\sqrt{1-\xi}c^Z_2M_Z^2\sqrt{\frac{\lambda^2}{4}+\frac{M_Z^2}{\hat{s}}}}{m_1^2\left(\lambda^2+\frac{12M_Z^2}{\hat{s}}\right)}+\mathcal{O}\left(\frac{v^4}{m_1^4}\right)\right],
\end{equation}
where\footnote{The coefficients $v_q$ and $a_q$ denote the vector and axial vector couplings of the $q\bar{q}$ pair with the $Z$ boson.} 
\begin{equation}
\label{analytic_2}
\sigma^{\rm SM}(\hat{s})=\frac{(g^{\rm SM}_Z)^2(v_q^2+a_q^2)\lambda\hat{s}\left(\lambda^2+\frac{12M_Z^2}{\hat{s}}\right)}{576\pi M_Z^2\left[(\hat{s}-M_Z^2)^2+M_Z^2\Gamma_Z^2\right]},\qquad 
\lambda=\sqrt{\left(1-\frac{M_Z^2}{\hat{s}}-\frac{m_h^2}{\hat{s}}\right)^2-\frac{4M_Z^2m_h^2}{\hat{s}^2}}\,.
\end{equation}
In the above equation the terms proportional to $c^Z_3$, $c^Z_4$, $c^Z_5$ are neglected in the limit where the quark masses are negligible.
In Fig.\,\ref{fig:zh_parton}, we display the cross section of the parton level process ($q\bar{q}\to ZH$) as a function of the partonic centre of mass energy $\sqrt{\hat{s}}$ for the elementary (red) and the composite Higgs setup with $\Gamma_{1,2}/m_{1,2}=1\%$ (blue) and $\Gamma_{1,2}/m_{1,2}=20\%$ (magenta,dashed). The simulated results with our implementation of the form factor given in Eq.\,\eqref{hvv_FF_3} are also plotted in the same figure and match well with the theoretical prediction for the cross section discussed above. 
The Fig.\,\ref{fig:zh_parton} shows that the presence of momentum dependent couplings at the $hZZ$ vertex in contrast to the elementary Higgs scenario causes an enhancement of the cross section near the tail of the distribution and eventually leads to a peak at $m_{1,2}$ due to the pole structure of Eq.\,\eqref{hvv_FF_3}.
However, we expect additional threshold contributions (depending on specific UV completion) near the $\sqrt{\hat{s}}\sim m_{1,2}$ that may significantly modify the cross section in this region. 
The grey dotted curve in the Fig.\,\ref{fig:zh_parton} denotes the cross section in the effective SILH framework as given in the Eq.\,\eqref{silh} with $c_H=1$, $c_W=c_B=1/2$ and $c_{HW}=c_{HB}=(8\pi^2f^2/m_1^2)\simeq12$.  
The dependence of the cross section on the parameters $c_{HW}$ and $c_{HB}$ are milder than the others. This choice of coefficients is adapted to match with the leading order expansion of our form factor parametrization and is well within the present limits (see Appendix\,\ref{SILH_map} for details).
In the region displayed in Fig.\,\ref{fig:zh_parton}, the SILH result matches fairly well with that obtained using the form factors when only the leading order terms in the expansion around $\hat{s}/m_1^2$ are kept in Eq.\,\eqref{analytic}. However, the cross-section at high centre of mass energies for the SILH case grows with energy as $\sigma\sim\hat{s}/m_\rho^4$, which would lead to a violation of tree level unitarity \cite{Biekotter:2014gup}. To correct for this, the SILH framework needs to be supplemented with higher derivative terms and additional dynamic degrees of freedom near the threshold region $\hat{s}\gtrsim m_\rho^2$. On the other hand, the cross section using the form factor decreases with $\hat{s}$ at large energies as $\sigma\sim M_Z^2/\hat{s}^2$ thereby preserving unitarity. Thus for the differential distributions, the form factor defined in Eq.\,\eqref{hvv_FF_3} may provide a more tractable resume of the composite Higgs in contrast with the elementary case.

In passing we note that the form factor scenario discussed here can in principle be distinguished from the $s$-channel exchange of a weakly coupled BSM particle like a $Z^\prime$ boson with $hZ^\prime Z^\prime$ coupling. 
Unlike the case for a $Z^\prime$, the form factor with proper normalization as given in the Eq.\,\eqref{hvv_FF_3} does not give any propagator suppression away from the pole of the mesonic states, rather it approaches 1.    
Further the interference between the $Z^\prime$ mediated process and the SM process creates differences in the distribution. There is discernible distortion in the line shape around the pole compared to the form factor case making them easily distinguishable.

We generate $10^5$ events at the leading order ({\tt LO}) and at centre of mass energy $\sqrt{s}=14$ TeV for both the elementary and the composite Higgs scenario in {\tt MadGraph5}, after applying the generator level cuts listed in Table~\ref{tab:gencuts}. These cuts have been chosen to remove soft $b$-jets and leptons and to ensure that the dilepton pair and the $b\bar{b}$ pair are produced from the final state objects. We parton shower the events using {\tt Pythia8} \cite{Sjostrand:2014zea}, jet-cluster using {\tt FastJet}\,\cite{Cacciari:2011ma}, pass them through detector simulation in {\tt Delphes} \cite{deFavereau:2013fsa} using the default {\tt CMS} card and finally generate the distributions in {\tt MadAnalysis5} \cite{Conte:2012fm}. 
\begin{table}[t]
	\centering
	\begin{tabular}{cccc}
		\hline\hline
		Observable & $p_T^{l,b}$ & $\eta_{l,b}$ & $\Delta R_{l^+l^-,b\bar{b}}$\\
		\hline
		Cut & $>$ 25 GeV & $<$ 2.5 & $<$ 3.0\\
		\hline\hline
	\end{tabular}
	\caption{\small\it Generator level cuts on the transverse momenta ($p^{l,b}_T$), pseudorapidity ($\eta_{l,b}$) and the angular separation ($\Delta R_{l^+l^-,b\bar{b}}$) of the final state objects, which are used to remove soft $b$-jets and leptons and to assure that the $l^+l^-$ and the $b\bar{b}$ pairs are produced from same outer legs, respectively.}
	\label{tab:gencuts}
\end{table}
\begin{figure}[t]
	\centering
	\subfloat[\label{fig:mll_initial}]{\includegraphics[scale=0.253]{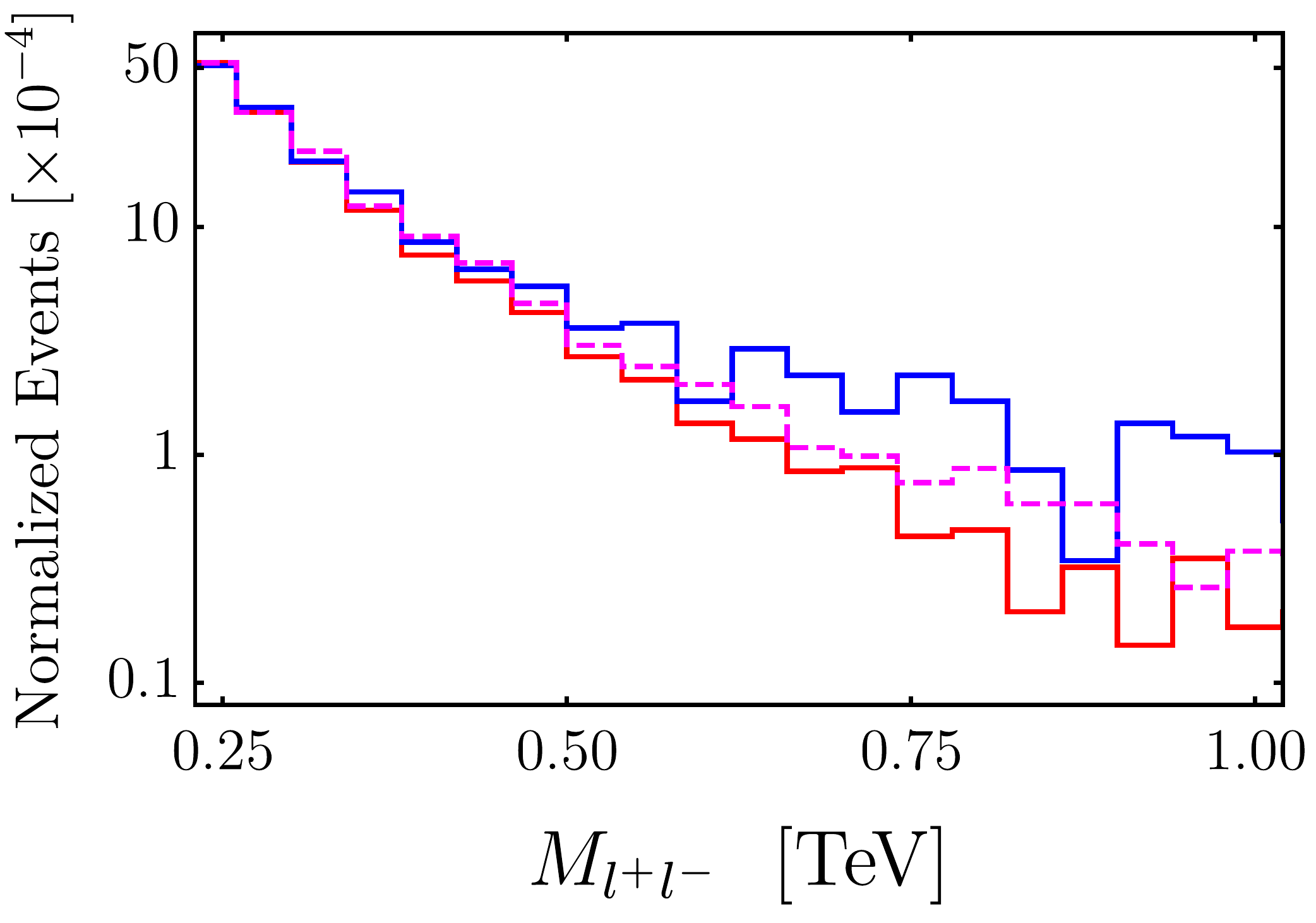}}
	\hspace{0.1cm}
	\subfloat[\label{fig:mllbb_initial}]{\includegraphics[scale=0.265]{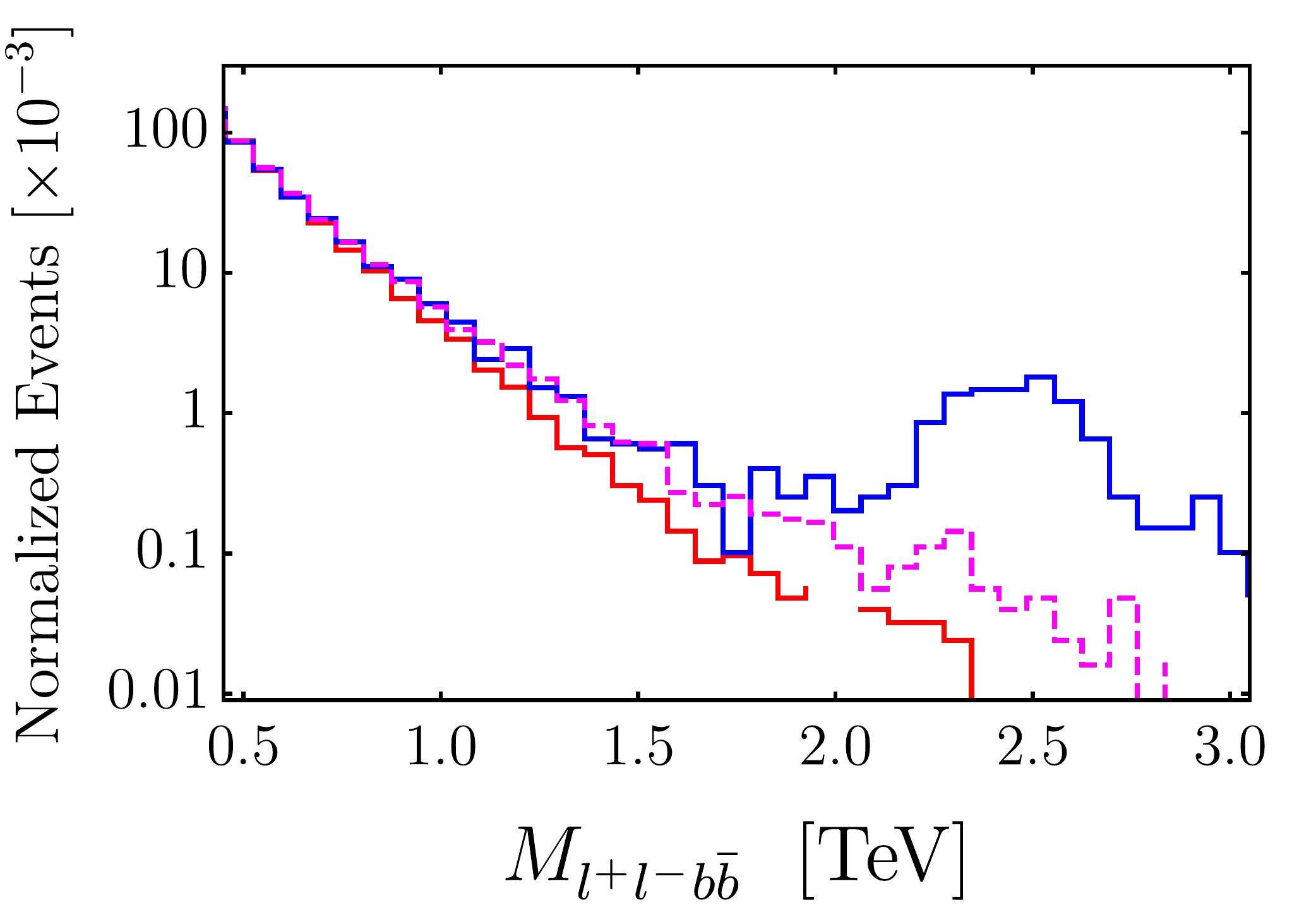}}
	\hspace{0.1cm}
	\subfloat[\label{fig:ptb_initial}]{\includegraphics[scale=0.253]{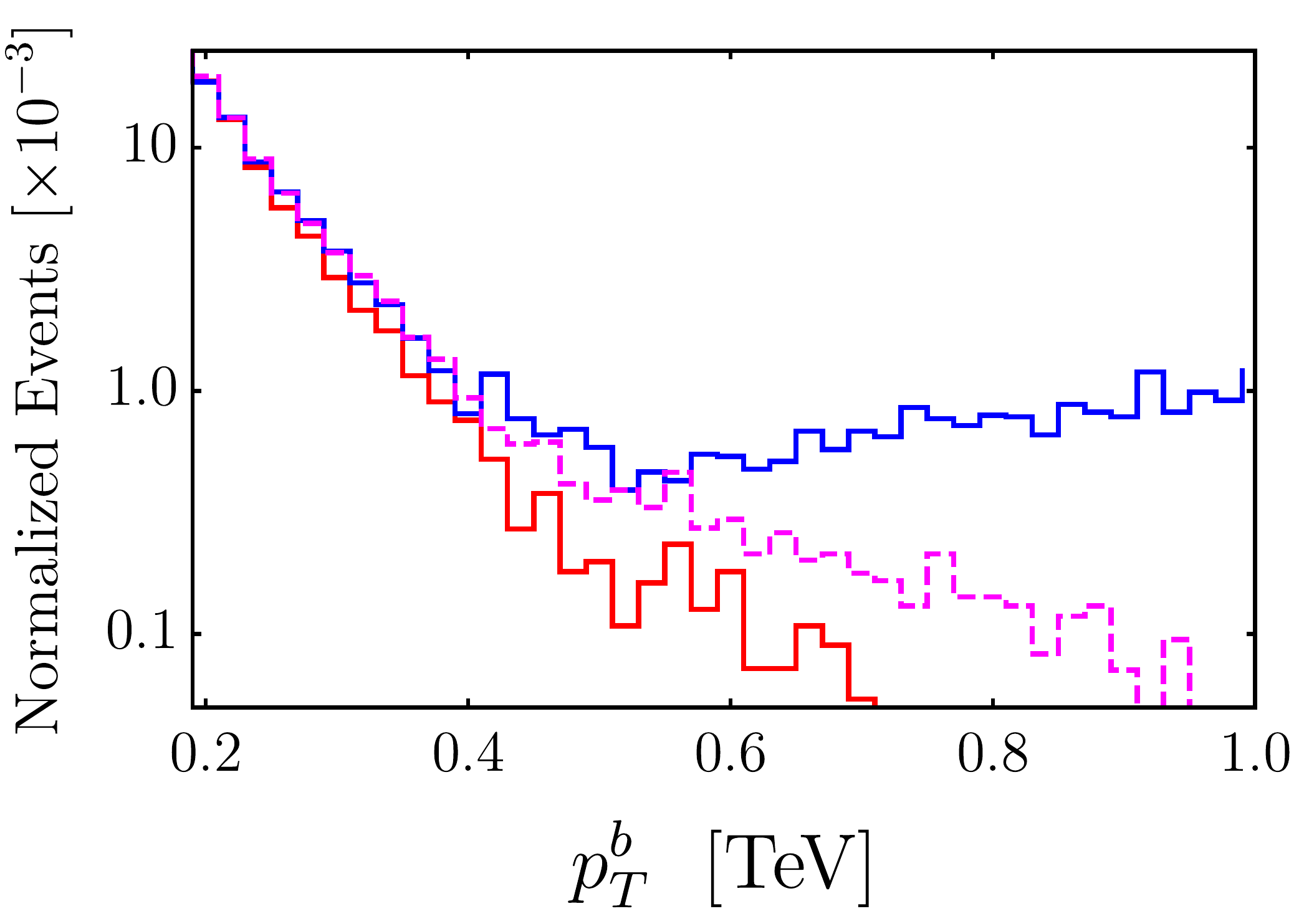}}\\
	\subfloat[\label{fig:ptl+_initial}]{\includegraphics[scale=0.253]{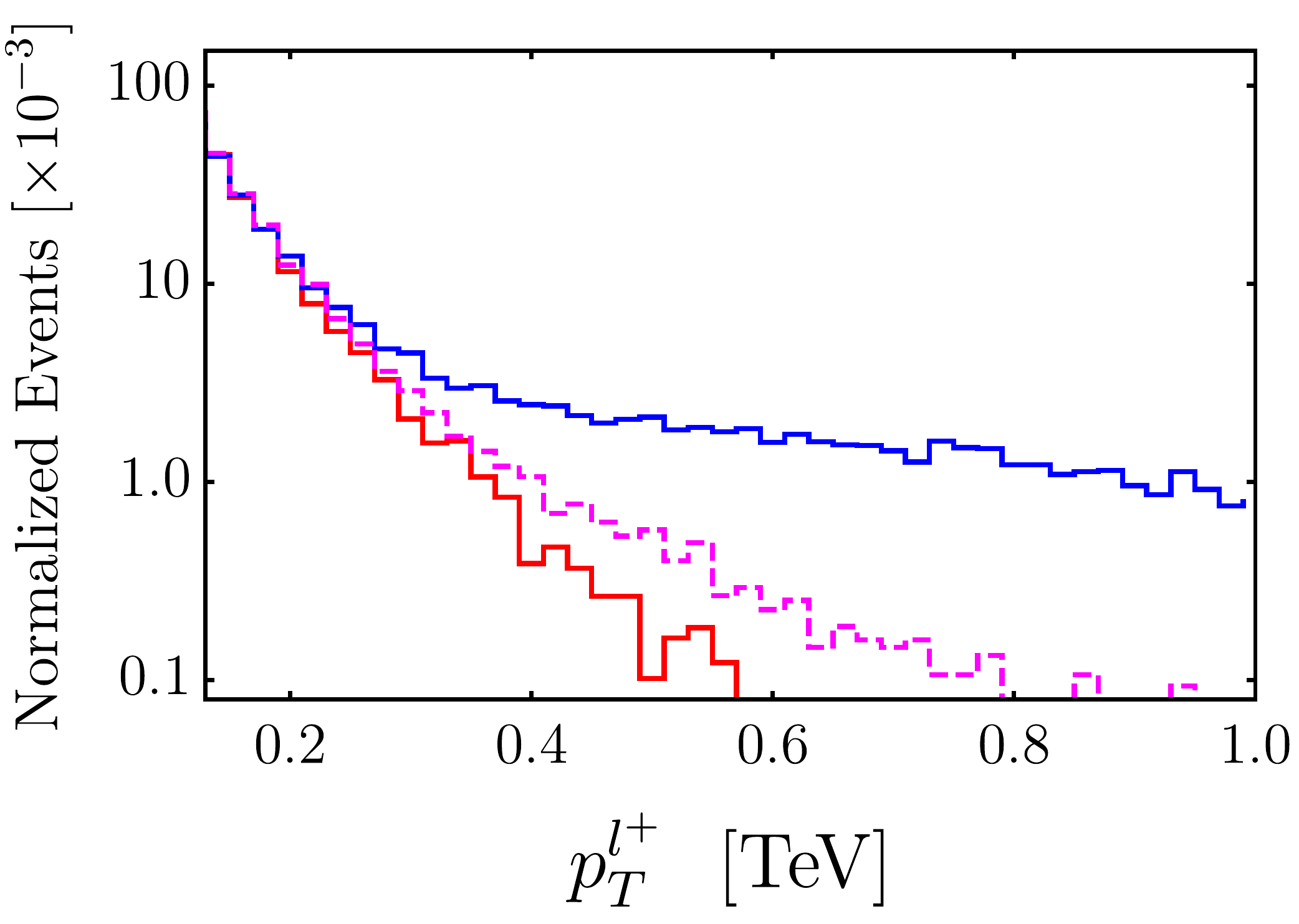}}
	\hspace{0.1cm}
	\subfloat[\label{fig:ptl-_initial}]{\includegraphics[scale=0.255]{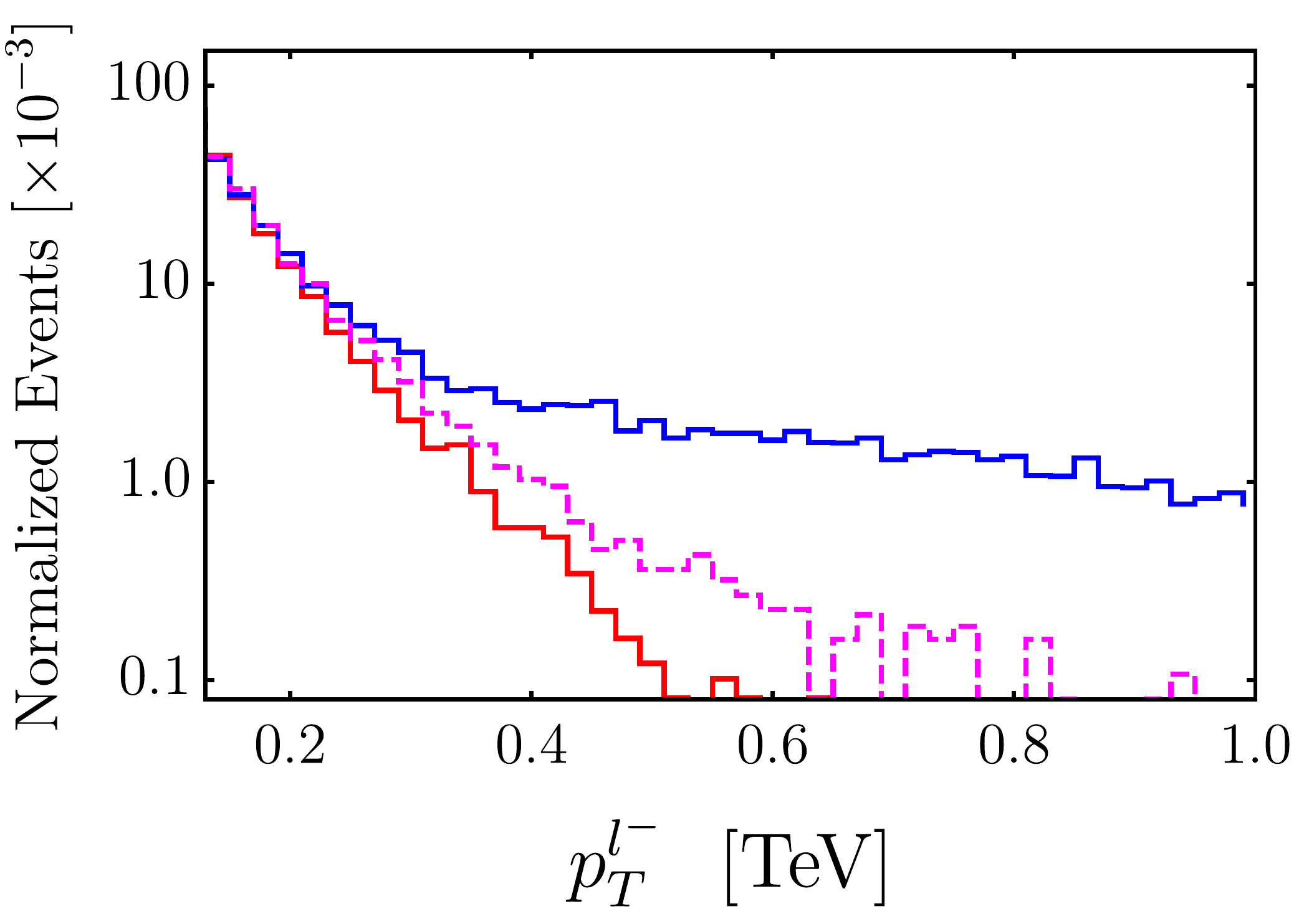}}
	\hspace{0.1cm}
	\subfloat[\label{fig:drll_initial}]{\includegraphics[scale=0.253]{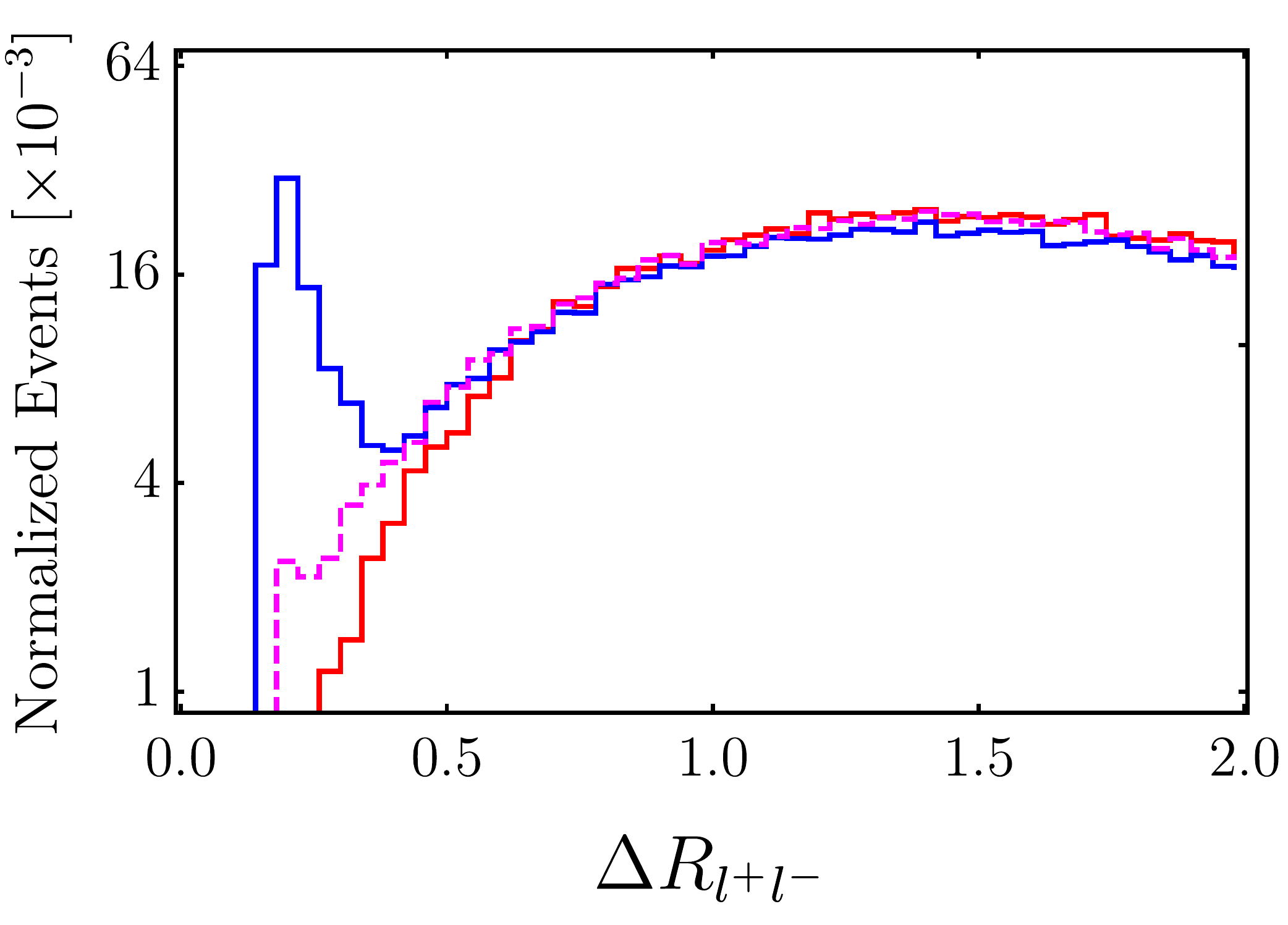}}
	\caption{\small\it Reconstructed level event distributions for observables displaying significant enhancement for the composite Higgs scenario with $\Gamma_{1,2}/m_{1,2}=1\%$ (blue) and $20\%$ (magenta,dashed) over the elementary case (red) near the tails. We choose $\xi=0.06$, $m_1=m_2=2.5$ TeV and $c_{i\ne 1}^{Z}=1$. The normalized events in the vertical axis represents the number of simulated events normalized to one event.}
	\label{fig:zh_reco_initial}
\end{figure}

In Fig.\,\ref{fig:zh_reco_initial}, we present the distributions of various kinematic parameters for the elementary (red) and the composite Higgs with $\Gamma_{1,2}/m_{1,2}=1\%$ (blue) and $20\%$ (magenta,dashed). 
In the vertical axis we plot the normalized events $N_i=\mathcal{N}_i/\mathcal{N}_{\rm tot}$, where $\mathcal{N}_i$  and $\mathcal{N}_{\rm tot}$ denote the number of events in the $i^{\rm th}$ bin and the total number of events for each scenario, respectively. 
We highlight the tails of the distributions where some deviations from the elementary case can be observed for all the observables presented. 
Leptons being one of the cleanest final state at the LHC, $p_T^{l^\pm}$ and $\Delta R_{l^+l^-}$ may provide an acceptable signal over the background, as discussed in the next section. The enhancement over the elementary scenario is pronounced when one of the momenta appearing in the form factor comes close to $m_1$ or $m_2$. Larger enhancement of the cross section at the tail is observed for the form factor involving narrower composite states. Evidently, with smaller values of $m_{1,2}$ more pronounced effects can be observed. 
In Fig\,\ref{fig:drll_initial} the bump for $\Gamma_{1,2}/m_{1,2}=1\%$ at small values of $\Delta R_{l^+l^-}$ corresponds to highly collimated final state leptons. The bump arises as a consequence of the threshold effects around $s\sim m_{1,2}^2$ due to the unique pole structure of the form factor in Eq.\,\eqref{hvv_FF_3}. The bump is suppressed for $\Gamma_{1/2}/m_{1/2}=20\%$ due to the diminished cross section around the threshold region for wide resonances. 
One interesting possibility is to investigate the hadronic decay of the $Z$ boson near the resonance, which has a larger branching ratio than the leptonic channel. The fat jets formed due to the collimation of the hadronic decay products of $Z$ near the pole of the form factor can be employed to efficiently discriminate the signal from the background using the jet substructure techniques \cite{CMS:2019emo}. However, we concentrate on the cleaner leptonic channel which may be a better discriminator away from the pole.
Our analysis emphasizes the importance of studying these differential distributions to probe a non-elementary nature of Higgs boson at the future runs of LHC.

\section{Reach at HL-LHC}
\label{sec:HLLHC}

\begin{figure}[t!]
	\centering
	\subfloat[\label{fig:mbb_final}]{\includegraphics[scale=0.38]{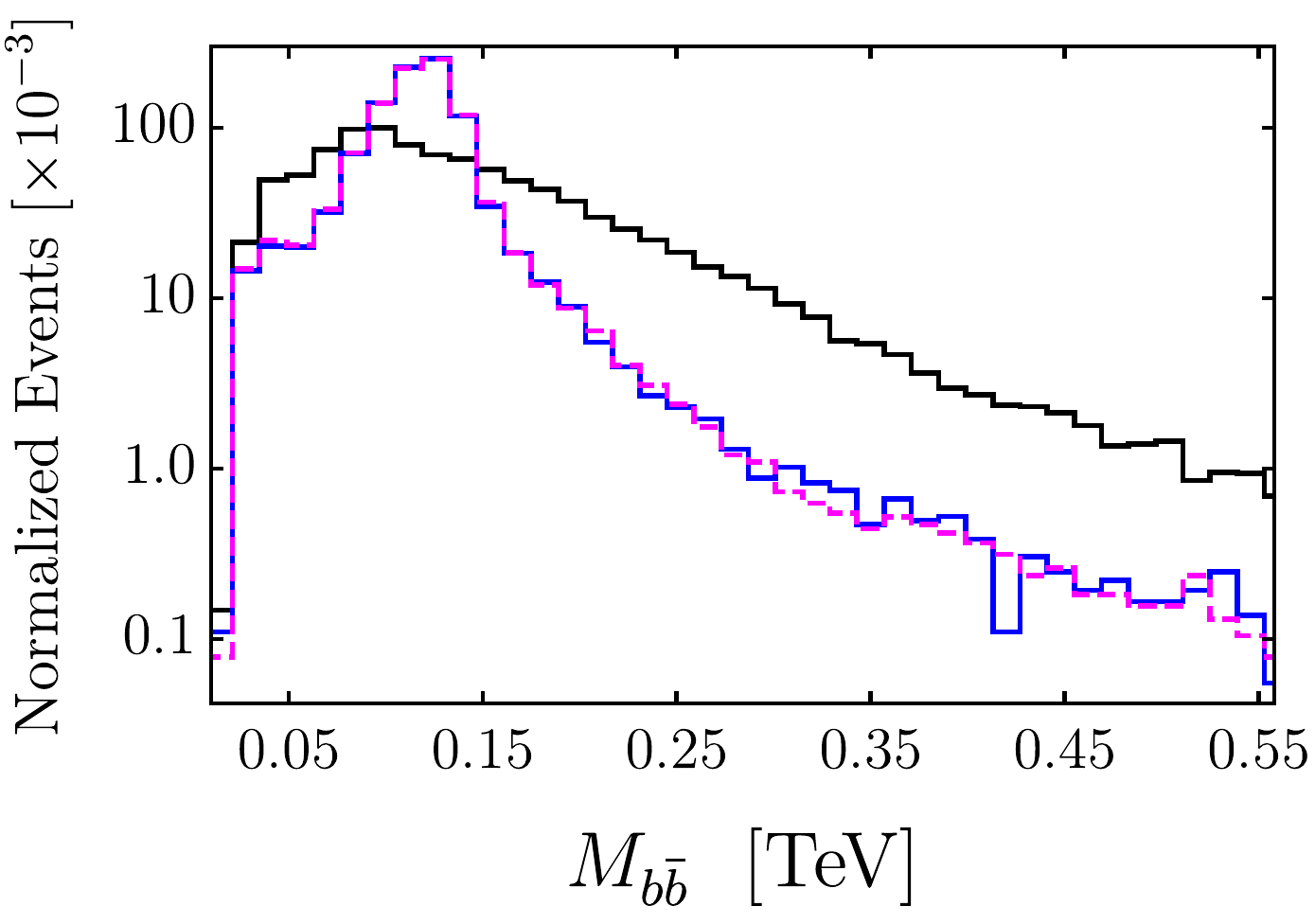}}
	\hspace{0.1cm}
	\subfloat[\label{fig:mll_final}]{\includegraphics[scale=0.38]{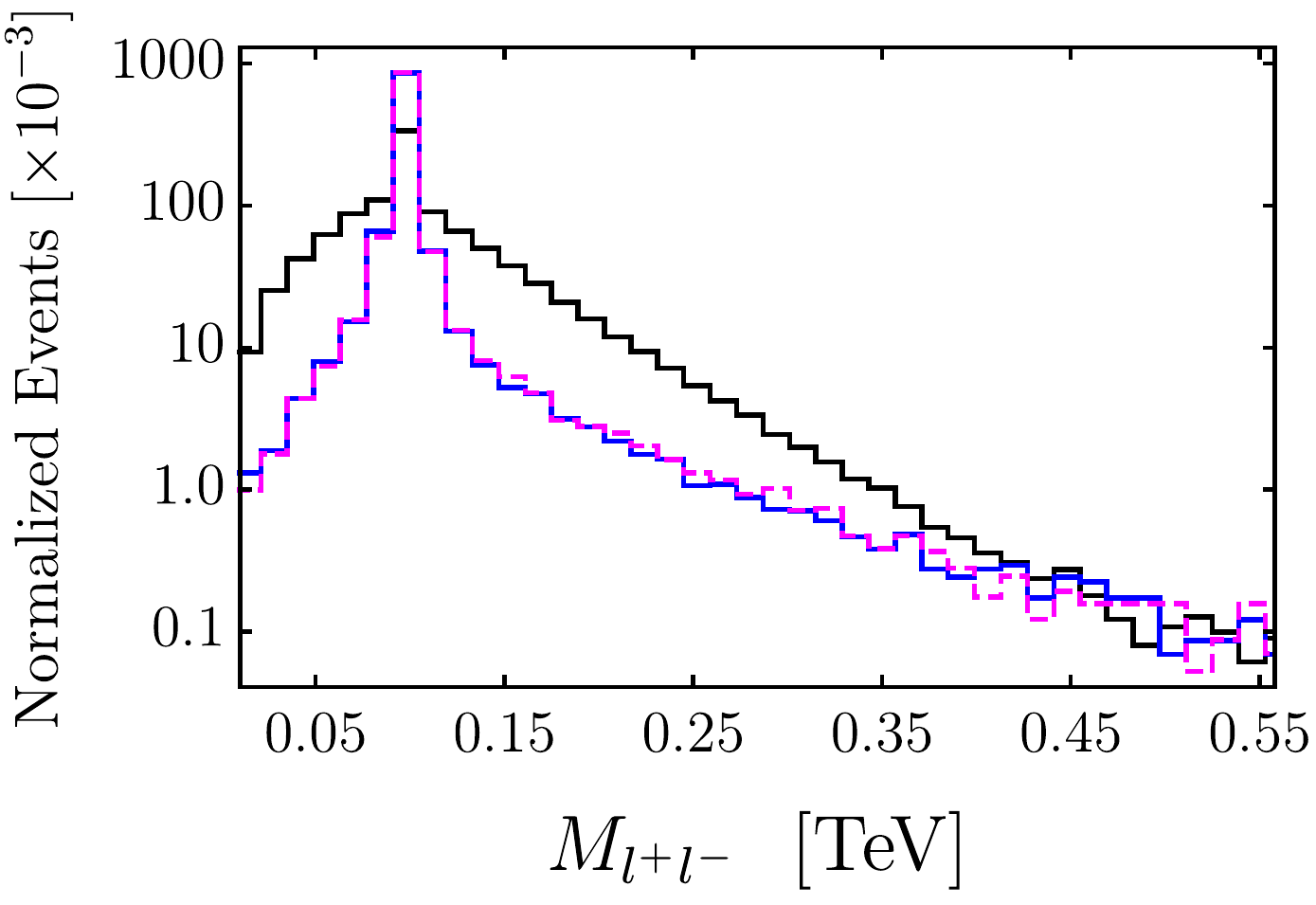}}
	\hspace{0.1cm}
	\subfloat[\label{fig:mllbb_final}]{\includegraphics[scale=0.38]{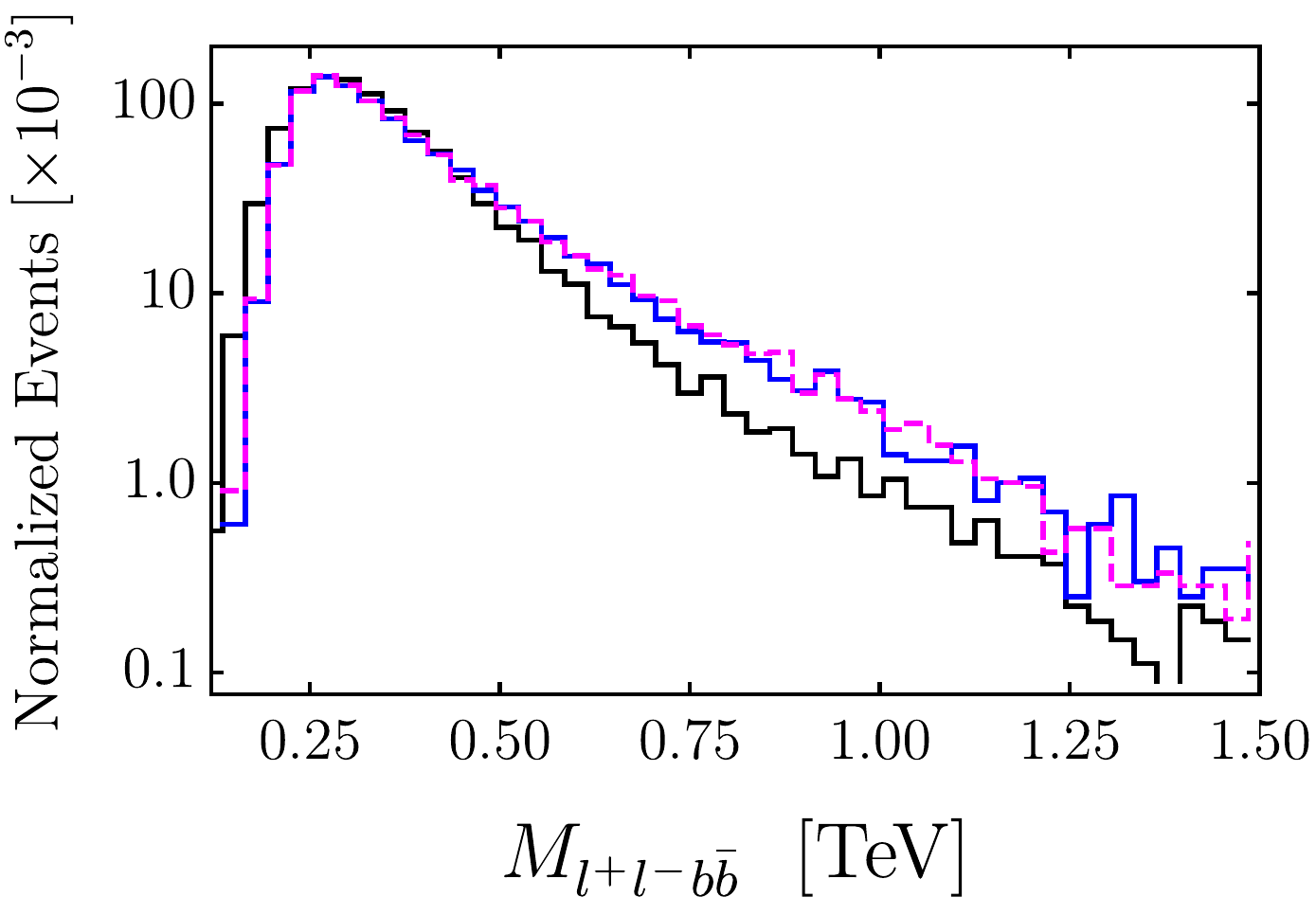}}\\
	\subfloat[\label{fig:ptb_final}]{\includegraphics[scale=0.38]{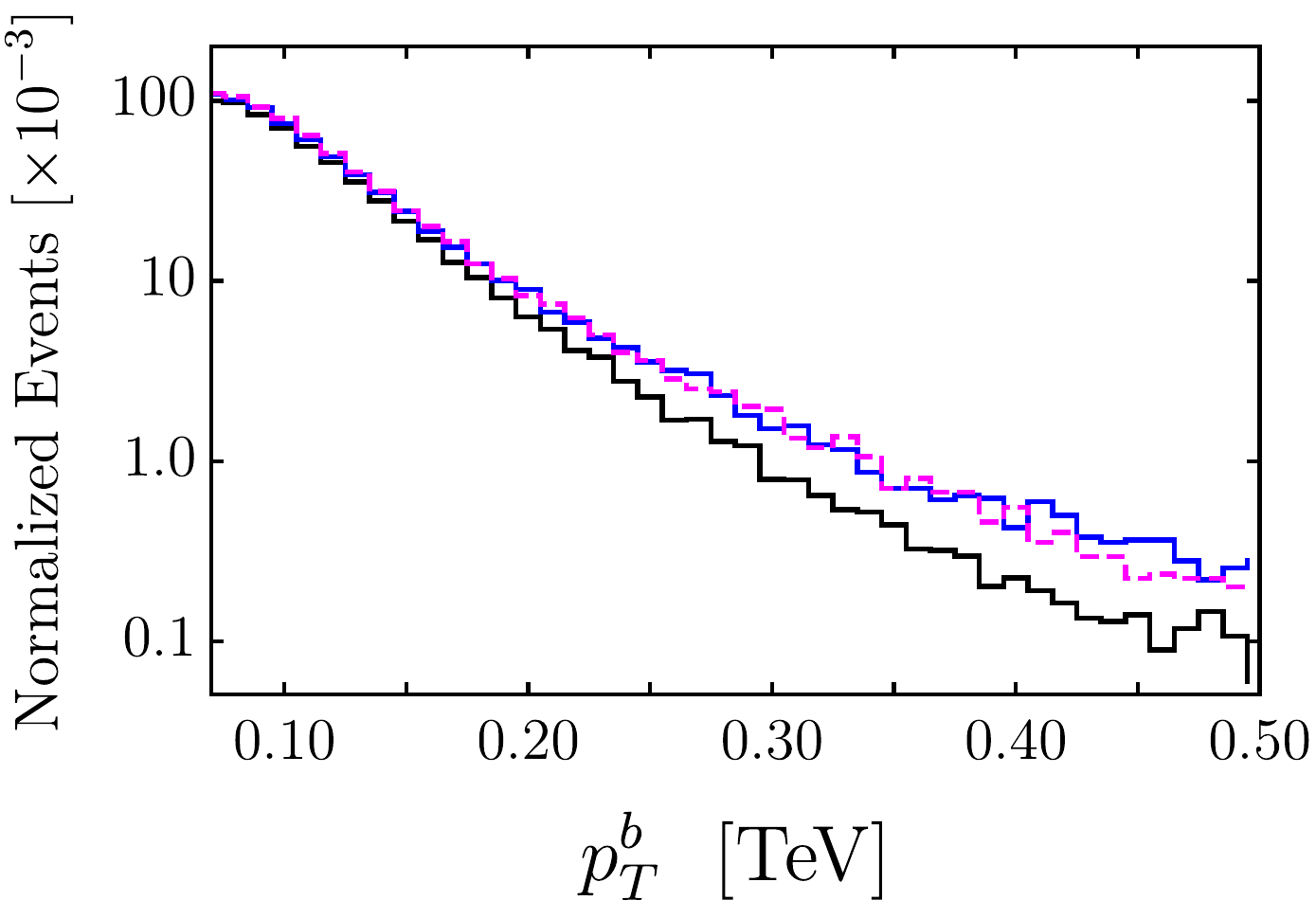}}
	\hspace{0.1cm}
	\subfloat[\label{fig:ptl+_final}]{\includegraphics[scale=0.38]{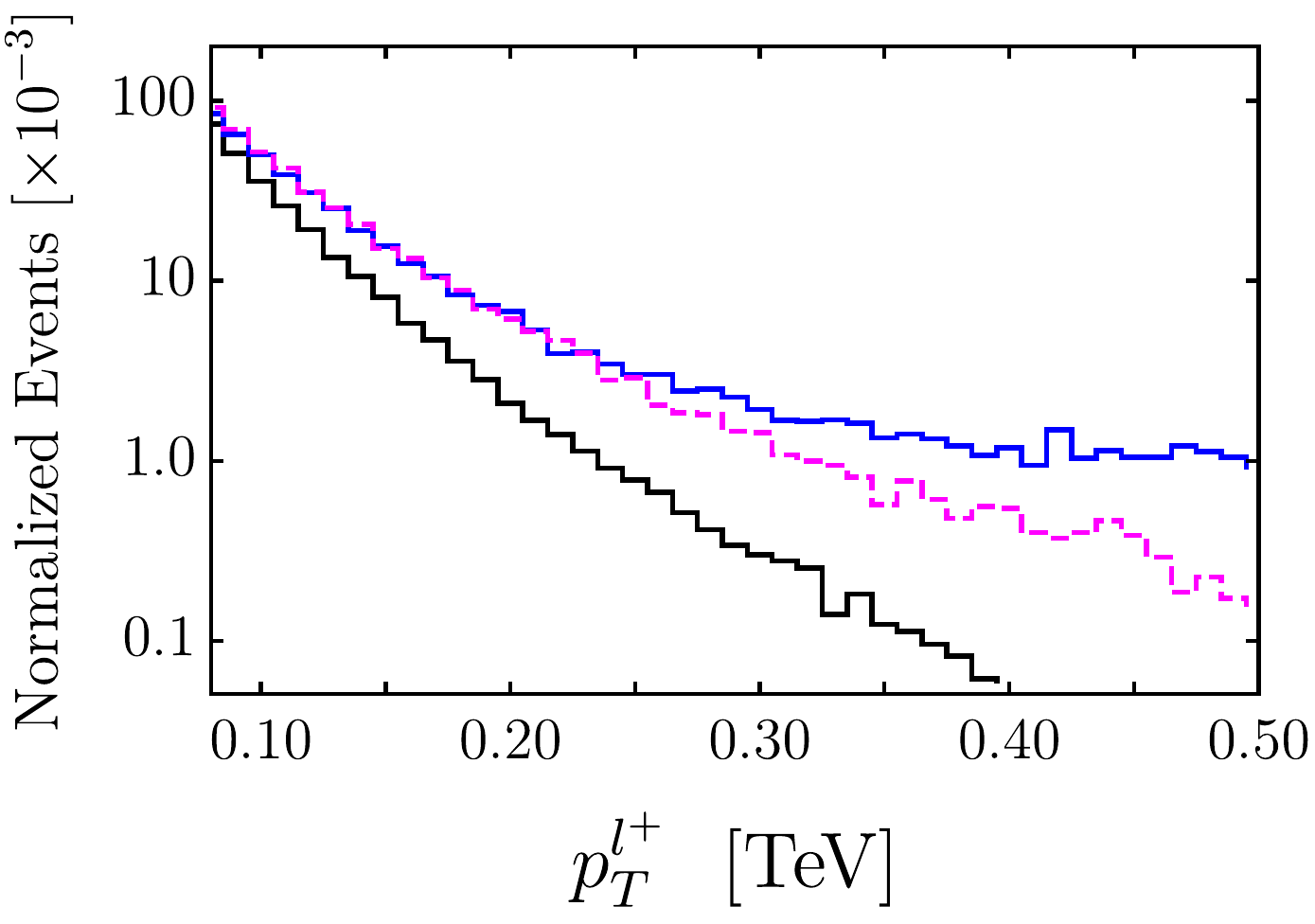}}
	\hspace{0.1cm}
	\subfloat[\label{fig:ptl-_final}]{\includegraphics[scale=0.38]{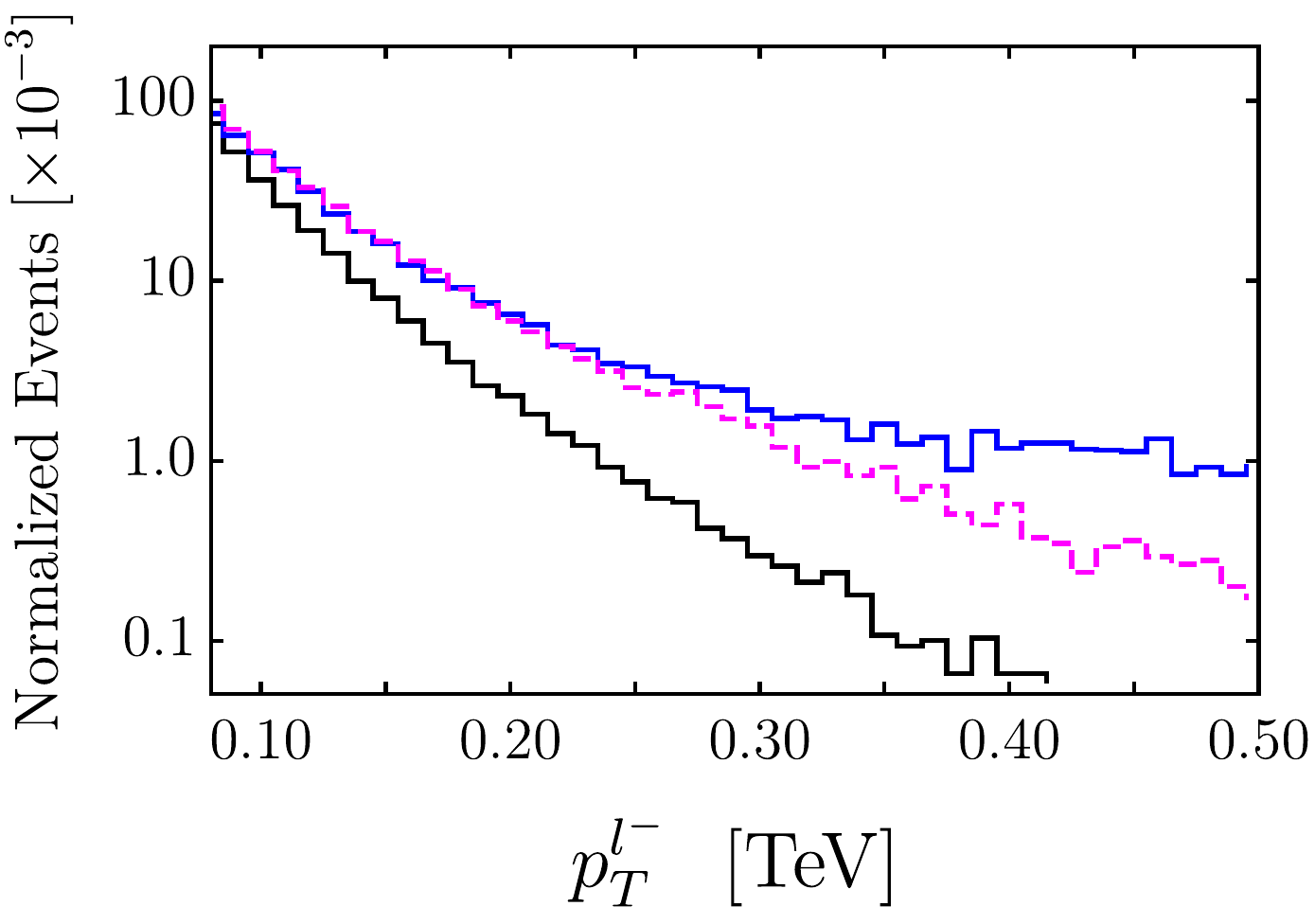}}\\
	\subfloat[\label{fig:deltarbb_final}]{\includegraphics[scale=0.38]{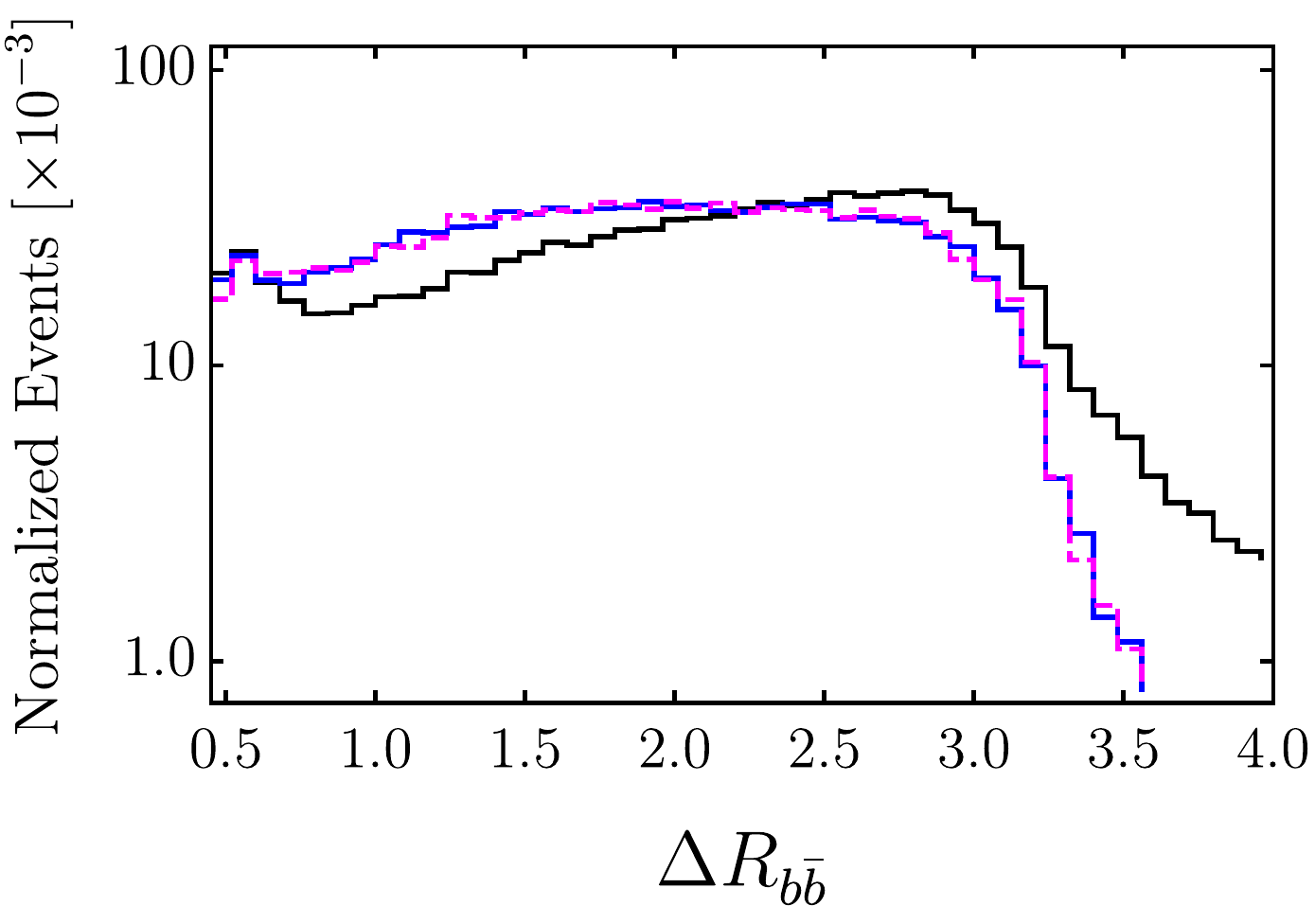}}
	\hspace{0.1cm}
	\subfloat[\label{fig:deltarll_final}]{\includegraphics[scale=0.38]{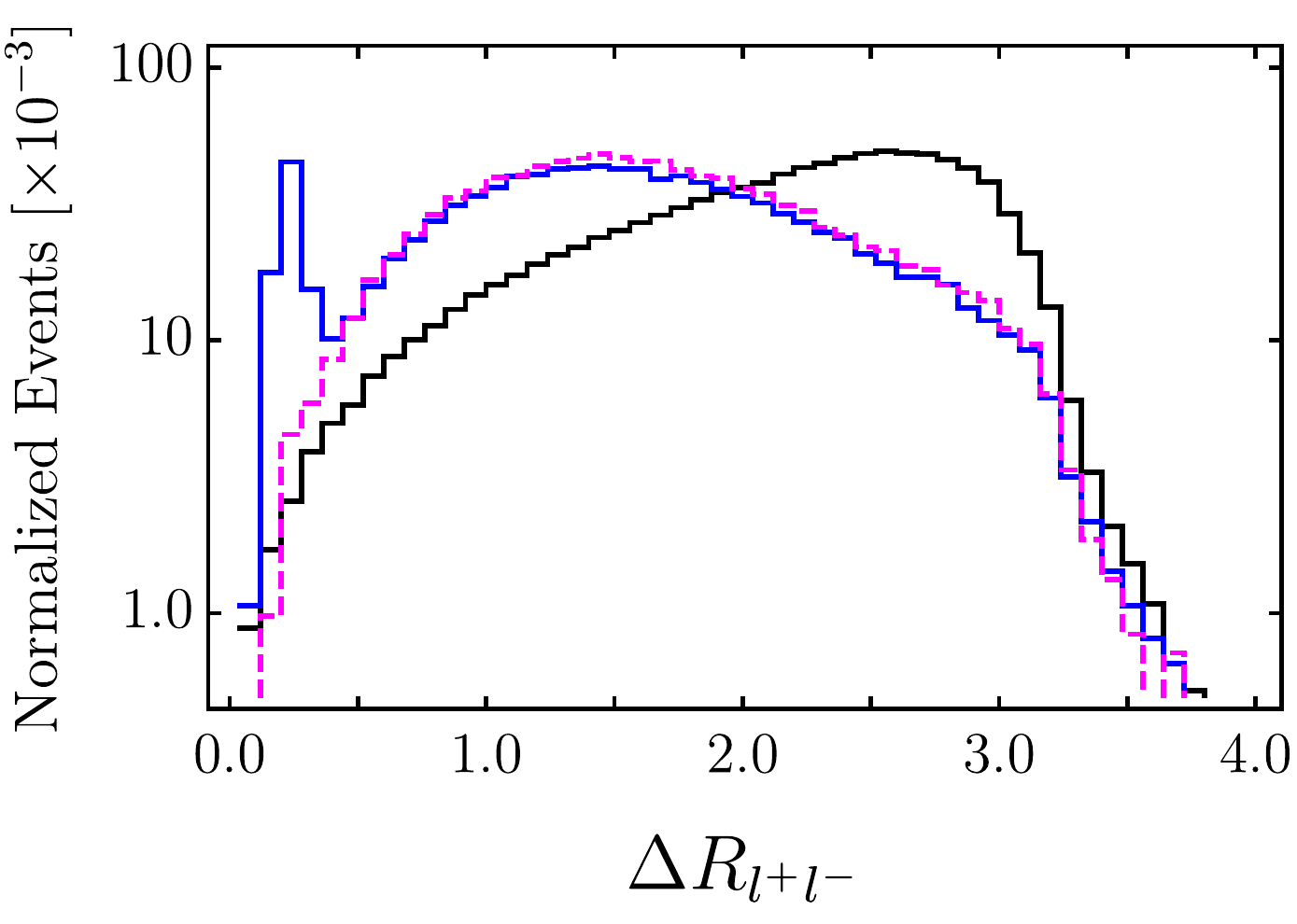}}
	\hspace{0.1cm}
	\subfloat[\label{fig:met_final}]{\includegraphics[scale=0.38]{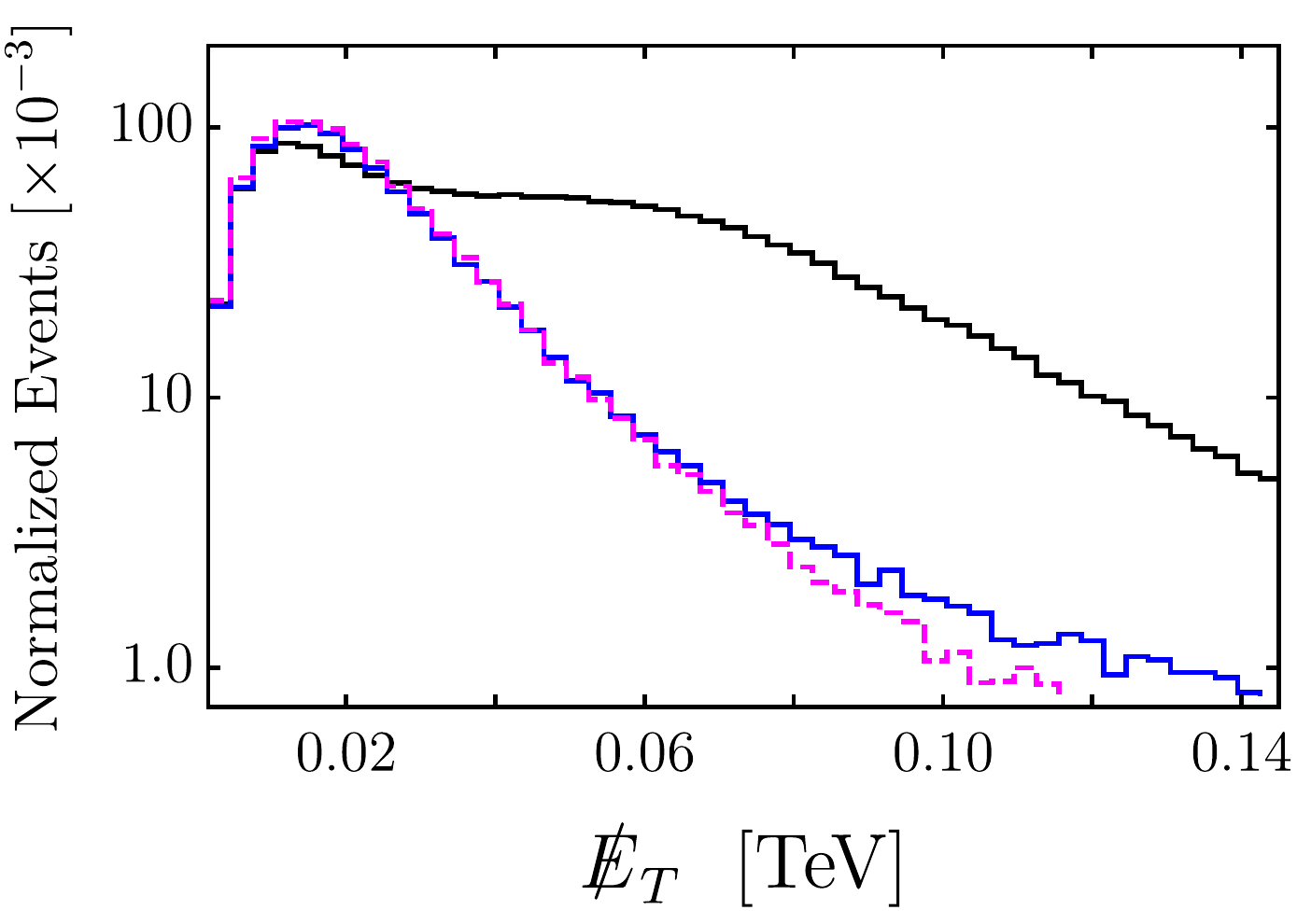}}
	\caption{\small\it Reconstructed level event distributions for the composite Higgs signal with $\Gamma_{1,2}/m_{1,2}=1\%$ (blue), 20\% (magenta,dashed) and the combined background events (black) are shown. The benchmark values of the model parameters are same as in Fig.\,\ref{fig:zh_reco_initial}.}
	\label{fig:zh_reco}
\end{figure}

\begin{figure}
	\centering
	\subfloat[\label{fig:significance1}]{\includegraphics[scale=0.7]{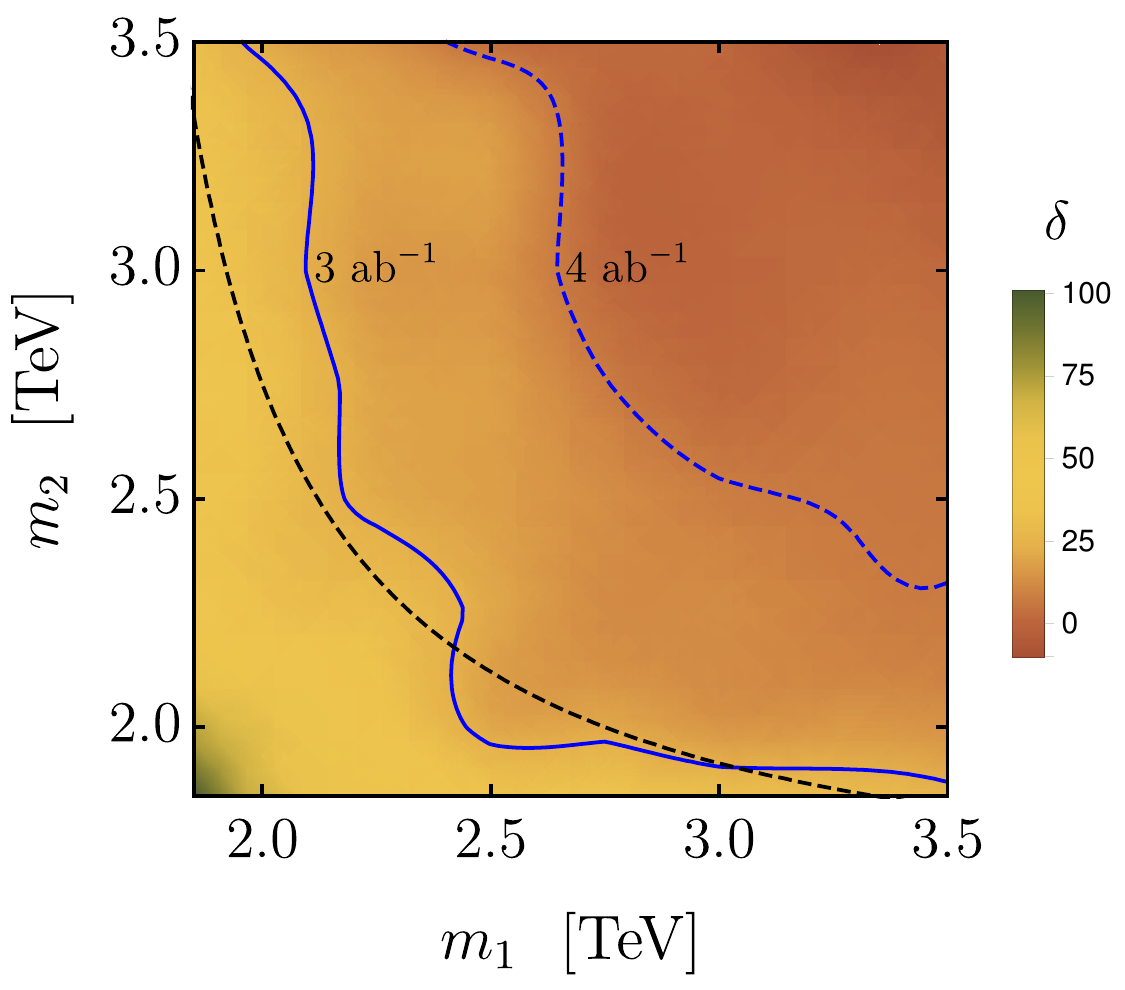}}\hspace{0.25cm}
	\subfloat[\label{fig:significance20}]{\includegraphics[scale=0.695]{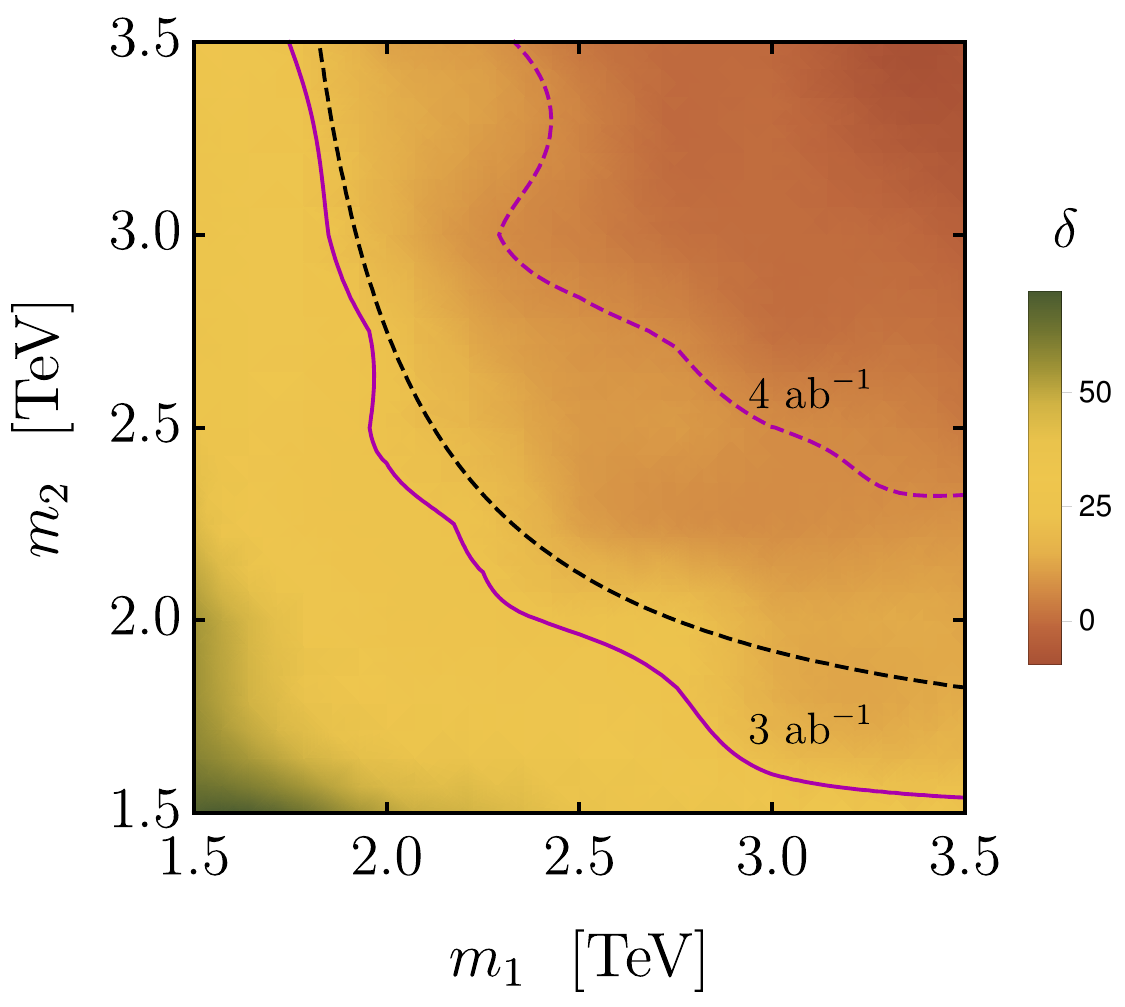}}
	\caption{\small\it Contours of $5\sigma$ signal significance in the $m_1-m_2$ parameter space at 3 $\rm ab^{-1}$ (solid) and 4 $\rm ab^{-1}$ (dashed) integrated luminosity are shown for $\Gamma_{1,2}/m_{1,2}=1\%$ (a) and 20\% (b), respectively. The density plot shows the deviation of number of expected signal events in the composite Higgs scenario in comparison to the elementary case. The black dashed lines in both the panels denote the limits from the $S$-parameter at 99\% CL \cite{Contino:2010rs}.}
	\label{fig:significance}
\end{figure}

\begin{figure}[t!]
	\centering
	\includegraphics[scale=0.65]{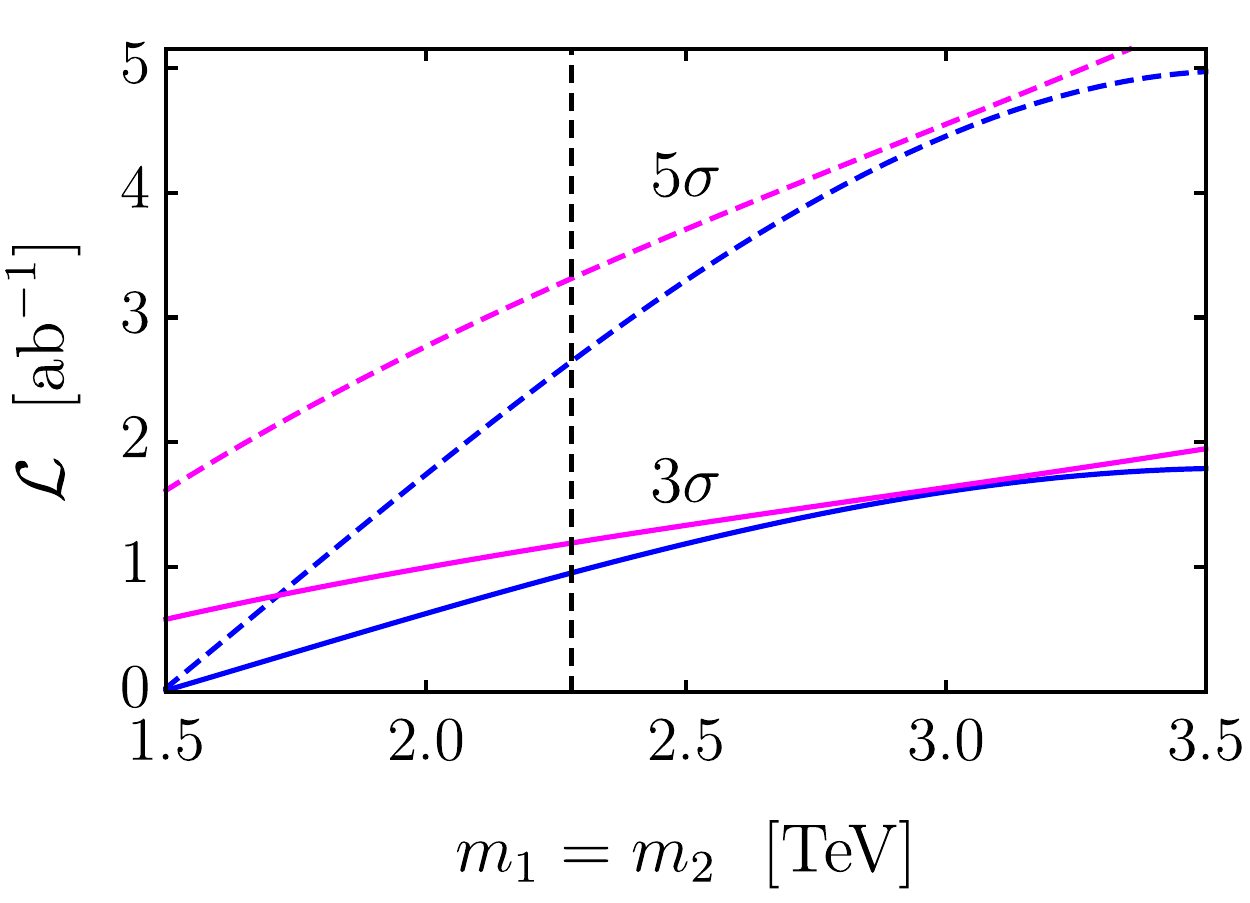}
	\caption{\small\it $5\sigma$ (dashed) and $3\sigma$ (solid) contours in the $(m_1=m_2)-\mathcal{L}$ plane for $\Gamma_{1,2}/m_{1,2}=1\%$ (blue) and $20\%$ (magenta). The black dashed line denotes the limit from the S-parameter at 99\% CL \cite{Contino:2010rs}.}
	\label{fig:luminosity}
\end{figure}

In this section we discuss the reach of HL-LHC to explore the parameter space of the composite Higgs couplings to $Z$-bosons in the associated Higgs production channel. The HL-LHC is expected to run at a centre of mass energy of 14 TeV and collect data till 3 $\rm ab^{-1}$ of integrated luminosity with the provision to reach up to 4 $\rm ab^{-1}$ \cite{Cepeda:2019klc}. 
The major backgrounds at the LHC from the signal channel considered here come from the $t\bar{t}$, single-top in the $tW$ channel, diboson and $Z$+jets events. We generate the background events by imposing the same generator level cuts given in the Table~\ref{tab:gencuts}. The {\tt LO} cross sections and efficiencies for the signal and background processes are displayed in the Table~\ref{tab:cross}. In Fig.\,\ref{fig:zh_reco}, we present the normalized kinematic distributions of various observables for both the background and the composite Higgs signals by generating $10^5$ events in each case. 
Taking cue from the distributions presented in Fig.\,\ref{fig:zh_reco} we devise the reconstructed level cuts as summarized in Table~\ref{tab:recocuts}. In the standard Higgs searches performed by CMS and ATLAS \cite{Sirunyan:2018kst,Aaboud:2018zhk}, cuts on $M_{l^+l^-}$ and $M_{b\bar{b}}$ are designed to focus on the Higgs peak providing a huge discrimination between the signal and the backgrounds. On the other hand we do not place any upper cut on the invariant momenta of the final states to focus on the tails of the distributions in order to maximize the deviation between the composite Higgs from the elementary Higgs scenario, at the cost of lowering overall signal to background significance.
To account for the higher order ({\tt NLO}) effects we multiply the {\tt LO} cross-sections with the appropriate $K$-factors for both the signal and the backgrounds, assuming that the efficiencies remain unchanged. In case of the composite Higgs signal we have used the same $K$-factor as obtained for the SM process which has been conservatively estimated as 1.2 \cite{Baglio:2020oqu}. The $K$-factors for the background are 1.47 for $t\bar{t}$, 1.09 for singletop in the $tW$ channel, 1.43 for diboson and 1.35 for $Z+$jets \cite{Alwall:2014hca,Stolarski:2020cvf}. 
We generate composite Higgs signal events by varying both $m_1$ and $m_2$ from $1.5-3.5$ TeV. The expected number of signal ($S$) and background ($B$) events at a certain integrated luminosity $\mathcal{L}$ is given by
\begin{equation}
\label{sig_def}
S,B=\sigma_{S,B}\times\epsilon_{S,B}\times\mathcal{L}\,,
\end{equation}
where $\sigma_{S,B}$ denote the signal and background cross sections obtained from {\tt MadGraph5} and multiplied with the appropriate $K$-factors, respectively. The signal (background) efficiencies $\epsilon_S$ ($\epsilon_{B}$), defined as the ratio of the number of signal (background) events surviving after applying the cuts listed in Table~\ref{tab:recocuts} to the initial number of events, are calculated for the elementary Higgs, the composite Higgs scenario and the background processes by generating $5\times 10^4$ ($1.5\times 10^5$ for backgrounds) events for each case. In Fig.\,\ref{fig:significance} we present the contours of $5\sigma$ signal significance (defined as $S/\sqrt{B}$) in the $m_1-m_2$ plane for $3~{\rm ab^{-1}}$ (solid) and $4~{\rm ab^{-1}}$ (dashed) integrated luminosities. 
In the same figure, the signal efficiency $\epsilon_S$ varies between $[0.013-0.036]$ for $\Gamma_{1,2}/m_{1,2}=1\%$ and  $[0.013-0.019]$ for $\Gamma_{1,2}/m_{1,2}=20\%$, respectively. 
Using the cuts given in Table~\ref{tab:recocuts} we find $S/\sqrt{B}$ for the elementary case are 3.92 and 4.53 at 3 $\rm ab^{-1}$ and 4 $\rm ab^{-1}$, respectively. 
\begin{table}[t]
	\centering
	\begin{tabular}{ccc}
		\hline\hline
		Signal & Cross-section (pb) & Efficiency\\
		\hline
		Composite Higgs (
		$\Gamma_{1,2}/m_{1,2}=1\%$) & $0.0109\pm 0.0003$ & $1.5\times 10^{-2}$\\
		Composite Higgs (
		$\Gamma_{1,2}/m_{1,2}=20\%$) & $0.0098\pm 0.0002$ & $1.5\times 10^{-2}$\\
		Elementary Higgs & $0.0100\pm 0.0002$ & $1.3\times 10^{-2}$\\
		\hline\hline
		Background & Cross-section (pb) & Efficiency\\
		\hline
		$t\bar{t} (\rightarrow l^+l^-+b\bar{b}+\slashed{E}_T)$ & $6.23\pm 0.01$ & $2.0\times 10^{-5}$\\
		$tW (\rightarrow l^+l^-+b+\slashed{E}_T)$ & $0.565\pm 0.005$ & $1.4\times 10^{-4}$\\
		diboson $ (\rightarrow l^+l^-+b\bar{b}+\slashed{E}_T)$ & $0.908\pm 0.001$ & $1.1\times 10^{-3}$ \\
		$Z+$jets $(\rightarrow l^+l^-+$jets$)$ & $287\pm 2$ & $1.1\times 10^{-5}$ \\
		\hline\hline
	\end{tabular}
	\caption{\small\it {\tt LO} cross sections and efficiencies for the signal (keeping $m_1=m_2=2.5$ TeV) and background processes at 14 TeV using our benchmark parameters and after applying the generator level cuts in {\tt MadGraph5}. 
	}
	\label{tab:cross}
\end{table}
To show the percentage deviation between the composite Higgs signal from the elementary case we define a quantity $\delta(m_1, m_2)$, as a function of $m_{1,2}$ as 
\begin{equation}
\delta(m_1,m_2)\equiv 100\times \frac{S_{\rm CH}(m_1,m_2)-S_{\rm EL}}{S_{\rm EL}}\,,
\end{equation}
where we calculate the expected number of signal events $S_{\rm CH}(m_1,m_2)$ for the composite Higgs scenario and $S_{\rm EL}$ for the elementary case using the Eq.\,\eqref{sig_def}. The density plot in Fig.\,\ref{fig:significance} shows the variation of $\delta(m_1,m_2)$ with the masses of the mesonic states. 
\begin{table}[t!]
	\centering
	\begin{tabular}{cccccccc}
		\hline\hline
		Observable & $p_T^b$ & $p_T^l$ & $M_{l^+l^-}$ & $M_{b\bar{b}}$ & $\Delta R_{l^+l^-,b\bar{b}}$ & $M_{l^+l^-b\bar{b}}$ &  $\slashed{E}_{T,miss}$ \\
		\hline
		Cut & $>$ 50 GeV & $>$ 50 GeV & $>$ 80 GeV  & $>$ 80 GeV & $<$ 2.0 & $>$ 500 GeV & $<$ 70 GeV \\
		\hline\hline 
	\end{tabular}
	\caption{\small\it Reconstructed level cuts used to focus on the tails of the distributions.}
	\label{tab:recocuts}
\end{table}
We find that the $\delta$ varies from 8\%--30\% (8\%--20\%) for $\Gamma_{1,2}/m_{1,2}=1\%(20\%)$ in the region explorable with at least $5\sigma$ signal significance at $4~{\rm ab}^{-1}$ integrated luminosity and allowed by the conservative limit from the $S$-parameter (shown by the black dashed line).
Note that for both cases the entire parameter space shown in the Fig.\,\ref{fig:significance} has a signal significance more than $3\sigma$. 
Inclusion of 1\% background systematic uncertainty can reduce the significance by up to 30\%. In Fig.\,\ref{fig:luminosity}, on the other hand, we show how the prospects of getting a $3\sigma$ ($5\sigma$) signal significance vary with the integrated luminosity for different values of $m_1(=m_2)$. We observe that for $\Gamma_{1,2}/m_{1,2}=1\%(20\%)$ and $\mathcal{L}\gtrsim2.6~{\rm ab}^{-1}(3.3~{\rm ab}^{-1})$, a significant region of allowed parameter space can be probed with 5$\sigma$ significance.

\section{Conclusions}
\label{sec:conclusion}

In this paper we demonstrate that a possible geometric shape of the Higgs boson may be probed at the future runs of the LHC, even if the compositeness scale is just beyond accessible range, by investigating the differential distributions of various Higgs production and decay channels. As a proof of principle we focus on the $pp\to Z^*H\to l^+l^-b\bar{b}$ channel in this paper. 

The couplings of composite pNGB Higgs boson with other SM particles are written in terms of momentum dependent form factors, which capture the essential features of the underlying strong dynamics. We construct the form factor involved in the three-point $hVV$ vertex from a bottom up approach, taking into account the results from large $N$ formalism. 
The collider analysis for the $pp\to Z^*H\to l^+l^-b\bar{b}$ channel shows that even below the reach of the resonance scales the composite Higgs signals start deviating from the elementary scenario as is evident from Fig.~\ref{fig:zh_reco_initial}. In this context we observe that for the $pp\to Z^*H\to l^+l^-b\bar{b}$ channel, the distributions of the total invariant mass of the final states, their individual $p_T$ and $\Delta R_{l^+l^-}$ may provide strong indications of a deviation from the elementary nature of the Higgs. It will be worthwhile to investigate the distributions of these (and other relevant) observables in more details at higher luminosity runs of the LHC. We further present the prospects for the HL-LHC to probe the Higgs form factor. Our conservative limits show that at $4~\rm ab^{-1}$ integrated luminosity, 5$\sigma$ signal significance can be achieved for a reasonable region of parameter space of the composite Higgs setup. 

\subsection*{Acknowledgements}

We thank Gabriele Ferretti and Diogo Buarque Franzosi for useful discussions. A.B. acknowledges support from the Knut and Alice Wallenberg foundation (Grant KAW 2017.0100, SHIFT project). A.B. was also supported by the Department of Atomic Energy, Govt. of India during the initial stage of the project. T.S.R. acknowledges Department of Science and Technology, Government of India, for support under Grant Agreement No. ECR/2018/002192 [Early Career Research Award]. S.D. acknowledges MHRD, Govt. of India for the research fellowship.

\appendix

\section{Mapping to the SILH framework}
\label{SILH_map}

To map the coefficients of the SILH Lagrangian with the form factor given in Eq.\,\eqref{hvv_FF_3}, we expand the latter upto leading order in $\mathcal{O}\left(p^2/m_\rho^2\right)$ as
\begin{align}
\label{hvv_FF_leading}
\Pi^{\mu\nu}_{V}\simeq f^2\left[\left(1-\frac{\xi}{2}+\frac{p_1^2+p_2^2}{m_1^2}\right)\eta^{\mu\nu} + \frac{1}{m_1^2}\big\{ c^V_2\left(\eta^{\mu\nu}p_1.p_2-p_2^\mu p_1^\nu\right) \right.
+ c^V_3 p_1^\mu p_2^\nu  + c^V_4 p_1^\mu p_1^\nu + c^V_5 p_2^\mu p_2^\nu \big\}\bigg]\,, 
\end{align}
\begin{table}[h!]
	\begin{center}
		\begin{tabular}{cc}
			\hline\hline
			SILH & Our parametrization \\
			\hline 
			$c_H$ & $1$\\
			$c_W+c_{HW}\left(\frac{m_\rho}{4\pi f}\right)^2$ & $1$\\
			$\cos^2\theta_W\left[c_W+c_{HW}\left(\frac{m_\rho}{4\pi f}\right)^2\right]+\sin^2\theta_W\left[c_B+c_{HB}\left(\frac{m_\rho}{4\pi f}\right)^2\right]$ & $1$\\
			$2c_{HW}\left(\frac{m_\rho}{4\pi f}\right)^2$ & $c^W_2$\\
			$2\left(\cos^2\theta_Wc_{HW}+\sin^2\theta_Wc_{HB}\right)\left(\frac{m_\rho}{4\pi f}\right)^2$ & $c^Z_2$\\
			$0$ & $c^W_3=c^Z_3$\\
			$c_W+c_{HW}\left(\frac{m_\rho}{4\pi f}\right)^2$ & $-c^W_4=-c^W_5$\\
			$\cos^2\theta_W\left[c_W+c_{HW}\left(\frac{m_\rho}{4\pi f}\right)^2\right]+\sin^2\theta_W\left[c_B+c_{HB}\left(\frac{m_\rho}{4\pi f}\right)^2\right]$ & $-c^Z_4=-c^Z_5$\\
			\hline\hline
		\end{tabular}
		\caption{\small\it Comparison between the coefficients of the SILH Lagrangian and the form factors.}
		\label{silh_mapping}
	\end{center}
\end{table}
where we have assumed $m_1=m_2=m_\rho$. In Table\,\ref{silh_mapping} we present the explicit expressions for the coefficients in Eq.\,\eqref{silh} for which the SILH Lagrangian can be mapped to $\Pi^{\mu\nu}_W$ and $\Pi^{\mu\nu}_Z$. 
In the Fig.\,\ref{fig:partonic}, we have employed values of the SILH coefficients using the Table\,\ref{silh_mapping} to match with the benchmark parameters chosen for the form factor parametrization. In particular, we assume $c_H=1$, $c_W=c_B=1/2$ and $c_{HW}=c_{HB}=(8\pi^2f^2/m_1^2)\simeq12$.
Note that the form factor approach strongly correlates the SILH coefficients.
\begin{table}[h!]
	\begin{center}
		\begin{tabular}{cccccccc}
			\hline\hline
			Coefficients & Relation with $c_i$ & Global fit \cite{Ellis:2018gqa} & ATLAS \cite{ATLAS:2019yhn} & CMS \cite{CMS:2020gsy} & Our choice \\
			\hline 
			$\bar{c}_H$ & $c_H\frac{v^2}{f^2}$ & $[-2.3, 0.1]$ & -- & -- & $0.06$ \\
			$\bar{c}_W+\bar{c}_B$ &	$(c_W+c_B)\frac{M_W^2}{m_\rho^2}$ & $[-0.05,0.05]$ & $0$ & $0$ & $0.001$ \\
			$\bar{c}_W-\bar{c}_B$ &	$(c_W-c_B)\frac{M_W^2}{m_\rho^2}$ & $[-0.12,0.04]$ & $[-0.006,0.014]$ & $[-0.09,0.08]$& $0$ \\
			$\bar{c}_{HW}$ & $c_{HW}\frac{M_W^2}{16\pi^2f^2}$ & $[-0.03,0.03]$ & $[-0.003,0.008]$ & $[-0.08,0.08]$ & $0.0005$\\
			$\bar{c}_{HB}$ & $c_{HB}\frac{M_W^2}{16\pi^2f^2}$ & $[-0.05,0.02]$ & $[-0.022,0.049]$& -- & $0.0005$\\
			\hline\hline
		\end{tabular}
		\caption{\small\it Limits at 95\% CL on the coefficients of the SILH Lagrangian. 
			In the last column we provide our choice of parameters with $f=1$ TeV and $m_\rho=m_1=m_2=2.5$ TeV.}
		\label{silh_limits}
	\end{center}
\end{table}
Bounds on the SILH coefficients are generally given in terms of the dimensionless parameters $\bar{c}_i\equiv c_i(M_W^2/\Lambda^2)$ \cite{Ellis:2018gqa,ATLAS:2019yhn,CMS:2020gsy,Ethier:2021bye}, where $\Lambda$ denotes the new physics scale. In Table\,\ref{silh_limits} current limits on the SILH coefficients and our choice of benchmark values are summarized. From the table it is evident that our choice of parameters is well within the present experimental bounds. We also list the explicit relations between $\bar{c}_i$ and the coefficients of each operators appearing in Eq.\,\eqref{silh}. 
\bibliographystyle{JHEP}
\bibliography{Higgs_FF}

\end{document}